\documentclass[11pt,a4paper]{article}%
\pdfoutput=1
\usepackage{jheppub}
\usepackage{epsfig}
\usepackage{hyperref}
\usepackage{amsmath,amssymb,graphicx,epstopdf,enumerate,amsfonts,booktabs,color}
\usepackage{amsmath}
\usepackage{amsfonts}
\usepackage{amssymb}
\usepackage{graphicx}%
\usepackage{relsize}
\usepackage{float}
\usepackage{placeins}
\usepackage[caption=false]{subfig}%

\providecommand{\U}[1]{\protect\rule{.1in}{.1in}}
\affiliation{Department of Electrophysics, National Chiao Tung University, Hsinchu, ROC}
\emailAdd{phyenjui@gmail.com}
\emailAdd{agoodmanjerry.ep02g@nctu.edu.tw}
\emailAdd{yiyang@mail.nctu.edu.tw}
\abstract{We study the holographic entanglement entropy in a $(d+1)$-dimensional
boundary quantum field theory at both the zero and finite temperature. The
phase diagrams for the holographic entanglement entropy at various
temperatures are obtained by solving the entangled surfaces in the different
homology. We also verify the Araki-Lieb inequality and illustrate the
entanglement plateau.}
\begin{document}

\title{Holographic Entanglement Entropy in Boundary Quantum Field Theory}
\author{En-Jui Chang, Chia-Jui Chou and Yi Yang}
\maketitle%

\setcounter{equation}{0}
\renewcommand{\theequation}{\arabic{section}.\arabic{equation}}%

\section{Introduction}

Entanglement is a pure quantum mechanics phenomenon inherent in quantum
states. It can be measured quantitatively by the entanglement entropy
associated with a specified entangled region $\mathcal{A}$ located on a Cauchy
slice of the spacetime by integrating out the degrees of freedom in its
complementary region $\mathcal{A}^{c}$. It has been shown that the leading
divergence term of the entanglement entropy is proportional to the area of the
entangling boundary, i.e. the boundary of the entangled region $\mathcal{A}$.
Furthermore, the finite part of the entanglement entropy contains non-trivial
information about the quantum states. It is well known that the entanglement
entropy of a given region and its complement are the same, $S_{\mathcal{A}%
}=S_{\mathcal{A}^{c}}$, for a system with only pure states. However, this is
not true anymore for a system with finite temperature. The difference $\delta
S_{\mathcal{A}}=S_{\mathcal{A}}-S_{\mathcal{A}^{c}}$ has been conjectured to
satisfy the Araki-Lieb inequality $\left\vert \delta S_{\mathcal{A}%
}\right\vert \leq S_{\mathcal{A\cup A}^{c}}$ \cite{AL}.

Calculating entanglement entropy is usually a not easy task in QFT.
Remarkably, by AdS/CFT correspondence \cite{9711200,9802150,9802109}, the
holographic entanglement entropy (HEE) was proposed to be the area of the
minimal entangled surface in \cite{0603001,0605073,0705.0016} and was
justified later on \cite{0606184,1006.0047,1102.0440,1304.4926}. This
prescription gives a very simple geometric picture to compute the HEE and has
been widely studied for the various holographic setups, for a review see
\cite{1609.01287}. It was shown that the Araki-Lieb inequality is due to the
homology constraint for the entangled surface
\cite{0704.3719,0710.2956,1305.3182}. For certain choices of the entangled
regions there are disconnected minimal surfaces satisfying the homology
constraint that leads to the famous phenomenon of the entanglement plateau
\cite{1306.4004}.

On the other hand, BQFT is a quantum field theory defined on a manifold with a
boundary where some suitable boundary conditions are imposed. It has important
applications in the physical systems with boundaries. For examples, string
theory with various branes and some condensed matter systems including Hall
effect, chiral magnetic effect, topology insulator etc. Several years ago,
holographic BQFT was proposed by extending the manifold where BQFT is defined
to a one-dimensional higher asymptotically AdS space, i.e. the bulk manifold,
with a geometric boundary \cite{1105.5165,1205.1573}. The key point of
holographic BQFT is thus to determine the shape of the geometric boundary in
the bulk. For the simple shapes with high symmetry such as the case of a disk
or half plane, many elegant results for BQFT have been obtained in
\cite{1105.5165,1205.1573,1108.5152}. Some interesting developments of BQFT
can be found in
\cite{1108.5152,1205.1573,1305.2334,1309.4523,1403.6475,1509.02160,1601.06418,1604.07571,1702.00566,1708.05080}%
.

Since both the entanglement entropy and BQFT can be studied holographically,
it is natural to investigate the HEE in holographic BQFT. The HEE in pure AdS
bulk spacetime has been studied in \cite{1701.04275,1701.07202,1703.04186}. It
was found that the proper boundary condition gives the orthogonal condition
that requires the minimal entangled surface must be normal to the geometric
boundaries if they intersect. The authors in \cite{1701.04275,1701.07202} also
found an interesting phenomenon that the entanglement entropy depends on the
distance to the boundary and carries a phase transition. In addition, the HEE
in the $(2+1)$-dimensional bulk manifold, such as $AdS_{3}$ and BTZ black
hole, has been considered in \cite{1410.7811,1511.03666}.

In this work, we study the HEE in a $(d+1)$-dimensional BQFT at both the zero
and finite temperatures. In the case of the zero temperature, we consider the
$(d+2)$-dimensional pure AdS spacetime as the bulk manifold. We find three
phases for the HEE depending on the size and the location of the entangled
region $\mathcal{A}$. Our result is consistent with the conclusion in
\cite{1701.04275,1701.07202}. In the case of the finite temperature, we
consider the $(d+2)$-dimensional Schwazschild-AdS black hole as the bulk
manifold. The similar three phases are found for the HEE. However, due to the
presence of the black hole horizon, the HEE for the entangled region
$\mathcal{A}$ and its complementary region $\mathcal{A}^{c}$ are different.
For BQFT at the finite temperature, entanglement entropy is mixed with the
thermal entropy given by the Bekenstein-Hawking entropy $S_{BH}$. We show that
the Araki-Lieb inequality $\left\vert \delta S_{\mathcal{A}}\right\vert \leq
S_{BH}$ always holds by a directly calculation. In addition, we obtain the
entanglement plateau for various sizes and locations of the entangled region
$\mathcal{A}$. Because there is a new phase due to the geometric boundary, the
entanglement plateau enjoys much richer structure in QFT with boundaries.

The paper is organized as follows. In section II, we briefly review the
holographic BQFT and present the solutions which we will use in this work. The
HEE in BQFT is calculated in section III. We discuss the phase structure of
the HEE at both the zero and finite temperatures. We also verify the
Araki-Lieb inequality and obtain the entanglement plateau. We summarize our
results in Section IV.%

\setcounter{equation}{0}
\renewcommand{\theequation}{\arabic{section}.\arabic{equation}}%

\section{Holographic Boundary Quantum Field Theory}

\begin{figure}[ptb]
\subfloat[Pure AdS spacetime]{
\includegraphics[scale=0.5]{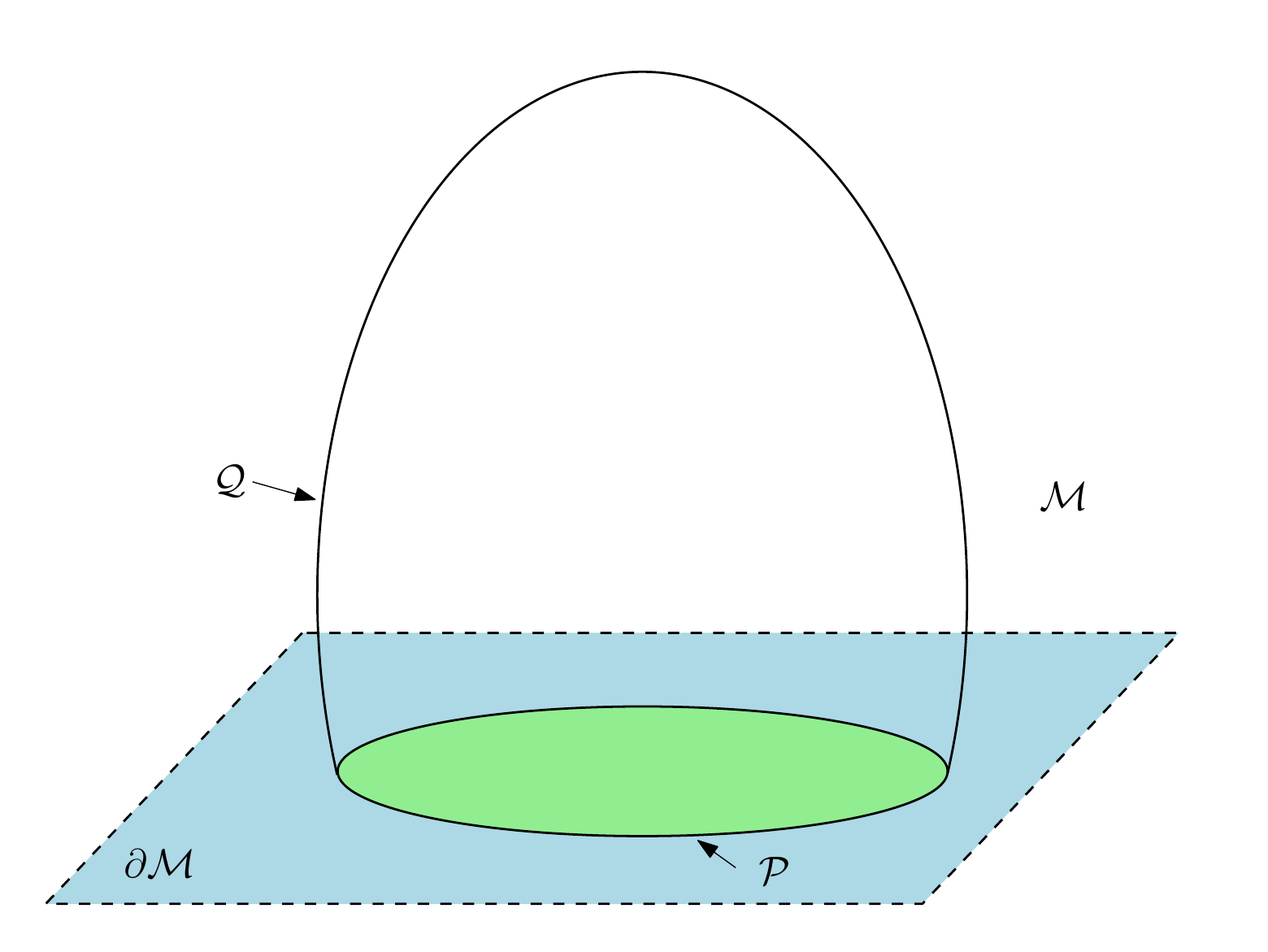} }
\subfloat[AdS black hole]{
\includegraphics[scale=0.5]{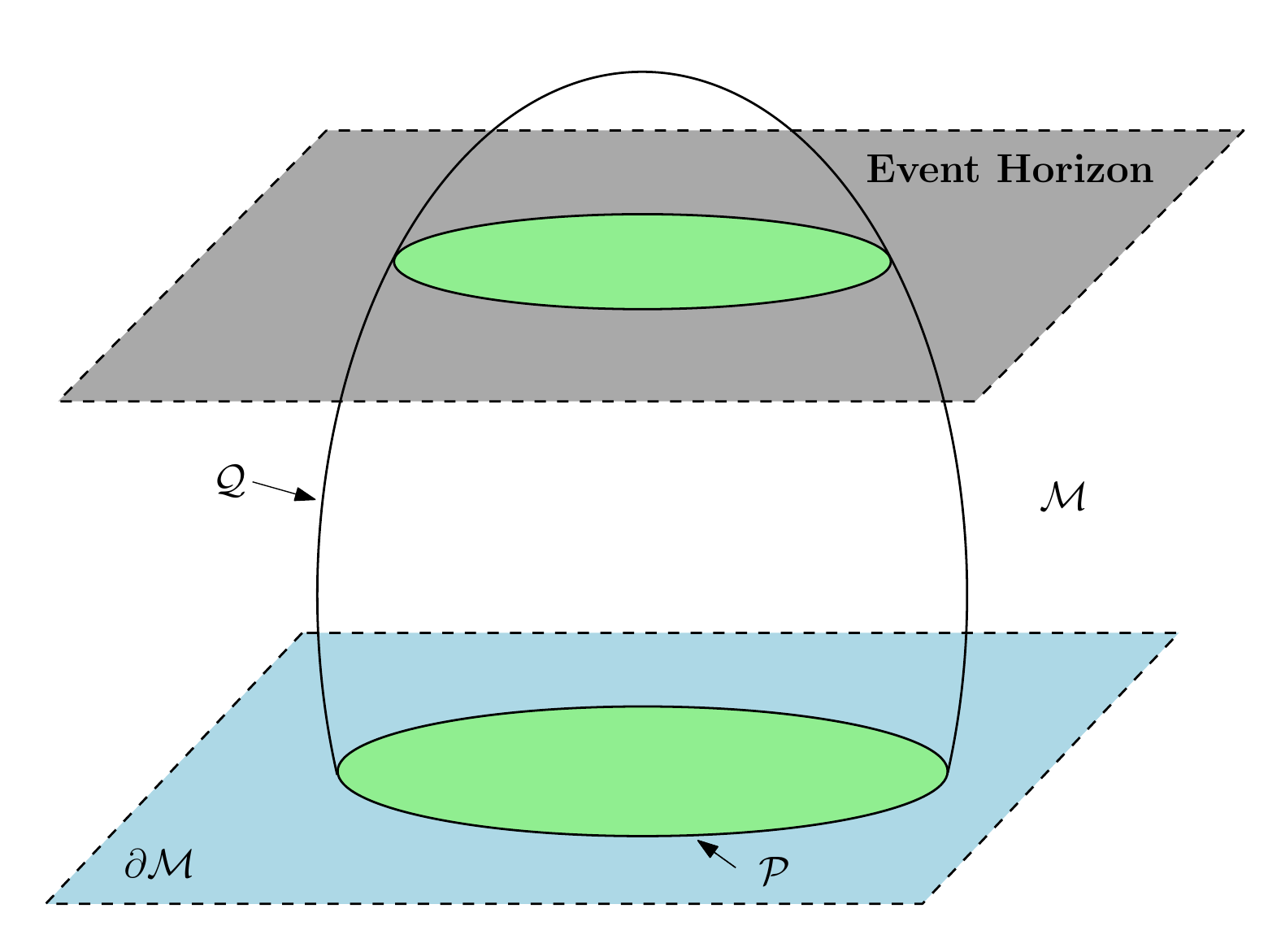} } \caption{Spacetime
setup for the holographic BQFT. (a) The bulk manifold is a pure AdS spacetime.
(b) The bulk manifold is an asymptotic AdS black hole with a horizon.}%
\label{Setup}%
\end{figure}

We consider a $(d+2)$-dimensional bulk manifold $\mathcal{M}$ which has a
$(d+1)$-dimensional conformal boundary $\partial\mathcal{M}$ as shown in the
Fig.\ref{Setup}. The bulk manifold $\mathcal{M}$ is either a pure AdS
spacetime, as in Fig.\ref{Setup}(a), or an asymptotic AdS black hole with an
event horizon, as in Fig.\ref{Setup}(b). In addition, there is a
$(d+1)$-dimensional hypersurface $\mathcal{Q}$ in $\mathcal{M}$ that
intersects the conformal boundary $\partial\mathcal{M}$ at a $d$-dimensional
hypersurface $\mathcal{P}$. A BQFT is defined on $\partial\mathcal{M}$ within
the boundary $\mathcal{P}$. The hypersurface $\mathcal{Q}$ could be considered
as the extension of the boundary $\mathcal{P}$ from $\partial\mathcal{M}$ into
the bulk $\mathcal{M}$ and represents a geometric boundary of the bulk. This
is our holographic setup for a BQFT living in $\partial\mathcal{M}$ with a
boundary $\mathcal{P}$.

The total action of the system is the sum of the actions of the various
geometric objects and their boundary terms,%
\begin{equation}
S=S_{\mathcal{M}}+S_{GH}+S_{\mathcal{Q}}+S_{\mathcal{P}},\label{action}%
\end{equation}
where
\begin{align}
S_{\mathcal{M}} &  =\int_{\mathcal{M}}\sqrt{-g}(R-2\Lambda_{\mathcal{M}}),\\
S_{\mathcal{Q}} &  =\int_{\mathcal{Q}}\sqrt{-h}(R_{\mathcal{Q}}-2\Lambda
_{\mathcal{Q}}+2K),\\
S_{\partial\mathcal{M}} &  =2\int_{\partial\mathcal{M}}\sqrt{-\gamma}%
K^{\prime},\\
S_{\mathcal{P}} &  =2\int_{\mathcal{P}}\sqrt{-\sigma}\theta,
\end{align}
and we have taken $16\pi G=1$. In the total action (\ref{action}),
$S_{\mathcal{M}}$ is the action of the bulk manifold $\mathcal{M}$ with $R$
and $\Lambda_{\mathcal{M}}$ being the intrinsic Ricci curvature and the
cosmological constant of $\mathcal{M}$. $S_{\mathcal{Q}}$ is the action of the
geometric boundary $\mathcal{Q}$ with $R_{\mathcal{Q}}$, $\Lambda
_{\mathcal{Q}}$ and $K$ being the intrinsic Ricci curvature, the cosmological
constant and the extrinsic curvatures of $\mathcal{Q}$ embedded in
$\mathcal{M}$. $S_{\partial\mathcal{M}}$ is the action of the conformal
boundary of $\partial\mathcal{M}$ with $K^{\prime}$ being the extrinsic
curvatures of $\partial\mathcal{M}$ embedded in $\mathcal{M}$. We remark that
the terms of $K$ and $K^{\prime}$ are the Gibbons-Hawking boundary terms for
the boundaries $\mathcal{Q}$ and $\partial\mathcal{M}$ of the bulk manifold
$\mathcal{M}$, respectively. Finally, $S_{\mathcal{P}}$ is the common boundary
term of $\mathcal{Q}$ and $\partial\mathcal{M}$ with $\theta=\cos
^{-1}{(n^{\mathcal{Q}}\cdot n^{\mathcal{M}})}$ being the supplementary angle
between $\mathcal{Q}$ and $\partial\mathcal{M}$, which makes a well-defined
variational principle on $\mathcal{P}$. Furthermore, $g_{ab}$ denotes the
metric of the bulk manifold $\mathcal{M}$, $h_{ab}=g_{ab}-n_{a}^{\mathcal{Q}%
}n_{b}^{\mathcal{Q}}$ and $\gamma_{ab}=g_{ab}-n_{a}^{\mathcal{M}}%
n_{b}^{\mathcal{M}}$ denote the induced metric of the boundaries $\mathcal{Q}$
and $\partial\mathcal{M}$, $\sigma_{ab}$ denotes the metric of $\mathcal{P}$.
Here we have defined the unit normal vectors of $\mathcal{Q}$ and
$\partial\mathcal{M}$ as $n^{\mathcal{Q}}$ and $n^{\mathcal{M}}$.

Varying $S_{\mathcal{M}}$ with $g^{ab}$ gives the equation of motion of the
bulk $\mathcal{M}$,
\begin{equation}
0=R_{ab}-\frac{1}{2}Rg_{ab}+\Lambda_{\mathcal{M}}g_{ab}.
\end{equation}
Varying $S_{\mathcal{Q}}$ with $h^{ab}$ gives the equation of motion of the
geometric boundary $\mathcal{Q}$,
\begin{equation}
{R_{\mathcal{Q}ab}}+2K_{ab}-\left(  \frac{1}{2}{R_{\mathcal{Q}}}%
+K-\Lambda_{\mathcal{Q}}\right)  h_{ab}=0,\label{NBC}%
\end{equation}
which is just the Neumann boundary condition originally proposed by Takayanagi
in \cite{1105.5165} and lately generalized by Chu et al. in \cite{1701.07202}
by adding the intrinsic curvature ${R_{\mathcal{Q}}}$.

The crucial problem in the construction of the holographic BQFT is to
determine the $(d+1)$-dimensional geometric boundary $\mathcal{Q}$ that
satisfies the boundary condition (\ref{NBC}). However, the boundary condition
(\ref{NBC}) is too strong to have a solution even in the pure AdS spacetime
because there are more constraint equations than the degrees of freedom. In
\cite{1701.04275,1701.07202}, the authors proposed the following mixed
boundary condition,
\begin{equation}
\left(  d-1\right)  \left(  {R_{\mathcal{Q}}}+2K\right)  -2\left(  d+1\right)
\Lambda_{\mathcal{Q}}=0.\label{BC}%
\end{equation}
Although it is still difficult to obtain a general solution of $\mathcal{Q}$
with the mixed boundary condition (\ref{BC}), it is possible to find solutions
in some special cases. A class of solutions satisfying the mixed boundary
condition (\ref{BC}) have been obtained in \cite{1701.04275,1701.07202}.

In this work, instead of constructing more solutions of the geometric boundary
$\mathcal{Q}$, our purpose is to study the boundary effect for the HEE in
BQFT. We will thus use an almost trivial solution of $\mathcal{Q}$ that is
perpendicular to the conformal boundary $\partial\mathcal{M}$ with a simple
embedding. Nevertheless, we will find rich phase structures of the HEE in our
simple geometry.

We present the solutions which will be used to study the HEE in detail in the following.

\subsection{Pure AdS}

To study a $(d+1)$-dimensional BQFT at the vanishing temperature, we consider
the bulk manifold $\mathcal{M}$ as the $(d+2)$-dimensional pure AdS spacetime
$AdS_{d+2}$ with the metric,
\begin{equation}
ds_{\mathcal{M}}^{2}=\frac{l_{AdS}^{2}}{z^{2}}\left(  -dt^{2}+dz^{2}
+\sum_{i=1}^{d}dx_{i}^{2} \right)  , \label{AdS}%
\end{equation}
where $l_{AdS}$ is the $AdS$ radius. The conformal boundary of $AdS_{d+2}$ is
a $(d+1)$-dimensional Minkowski spacetime located at $z=0$.

We propose a simple solution of the geometric boundary $\mathcal{Q}$ as a
$(d+1)$-dimensional hepersurface embedded in the bulk manifold as,
\begin{equation}
ds_{\mathcal{Q}}^{2}=\frac{l_{AdS}^{2}}{z^{2}}\left(  -dt^{2}+dz^{2}%
+\sum_{i=2}^{d}dx_{i}^{2}\right)  ,\label{AdS-Q}%
\end{equation}
with a simple embedding $x_{1}=$ constant. The intrinsic curvature, the
extrinsic curvature and the cosmological constant on $\mathcal{Q}$ can be
calculated as,%
\begin{equation}
R_{\mathcal{Q}}=-\frac{d(d+1)}{l_{AdS}^{2}}\text{, }K_{ab}=0\text{, }%
\Lambda_{\mathcal{Q}}=-\frac{d(d-1)}{2l_{AdS}^{2}}.
\end{equation}
It is easy to verify that the mixed boundary condition (\ref{BC}) is satisfied.

This simple solution is a special case of the solutions constructed in
\cite{1701.04275,1701.07202} with $\theta=0$, i.e. the geometric boundary
$\mathcal{Q}$ is perpendicular to $\partial\mathcal{M}$ at their intersection
$\mathcal{P}$.

\subsection{Schwarzschild-AdS Black Hole}

To study the $(d+1)$-dimensional BQFT at a finite temperature, we consider the
bulk manifold $\mathcal{M}$ as the $(d+2)$-dimensional Schwarzschild-AdS black
hole with the metric,%
\begin{equation}
ds_{\mathcal{M}}^{2}=\frac{l_{AdS}^{2}}{z^{2}}\left(  -g\left(  z\right)
dt^{2}+\frac{dz^{2}}{g\left(  z\right)  }+\sum_{i=1}^{d}dx_{i}^{2}\right)
,\label{AdS-Schw}%
\end{equation}
with
\begin{equation}
g\left(  z\right)  =1-\frac{z^{d+1}}{z_{H}^{d+1}},
\end{equation}
The Hawking temperature and the Bekenstein-Hawking entropy density of the
$(d+2)$-dimensional Schwarzschild-AdS black hole are
\begin{equation}
T=\frac{d+1}{4\pi z_{H}}\text{, }S_{BH}=\frac{l_{AdS}^{d}L^{d}}{4z_{H}^{d}%
}.\label{T and S}%
\end{equation}
Similar to the pure AdS case, we propose a solution of the geometric boundary
$\mathcal{Q}$ as a $(d+1)$-dimensional hepersurface embedded in the bulk
manifold as,
\begin{equation}
ds_{\mathcal{Q}}^{2}=\frac{l_{AdS}^{2}}{z^{2}}\left(  -g\left(  z\right)
dt^{2}+\frac{dz^{2}}{g\left(  z\right)  }+\sum_{i=2}^{d}dx_{i}^{2}\right)
,\label{AdS-Schw-Q}%
\end{equation}
with a simple embedding $x_{1}=$ constant. The intrinsic curvature, the
extrinsic curvature and the cosmological constant on $\mathcal{Q}$ are the
same as the pure AdS case,%
\begin{equation}
R_{\mathcal{Q}}=-\frac{d(d+1)}{l_{AdS}^{2}}\text{, }K_{ab}=0\text{, }%
\Lambda_{\mathcal{Q}}=-\frac{d(d-1)}{2l_{AdS}^{2}},
\end{equation}
which satisfy the mixed boundary condition (\ref{BC}).%

\setcounter{equation}{0}
\renewcommand{\theequation}{\arabic{section}.\arabic{equation}}%

\section{Holographic Entanglement Entropy}

Without the geometric boundary $\mathcal{Q}$, the prescription to compute the
entanglement entropy holographically for the static situations has been
addressed by Ryu and Takayanagi (RT) in \cite{0603001,0605073}. It was later
generalized in \cite{0705.0016}\ to general states including arbitrary time
dependence. In this work, we focus on the static situation.

\begin{figure}[ptb]
\subfloat[Pure AdS spacetime]{
\includegraphics[scale=0.5]{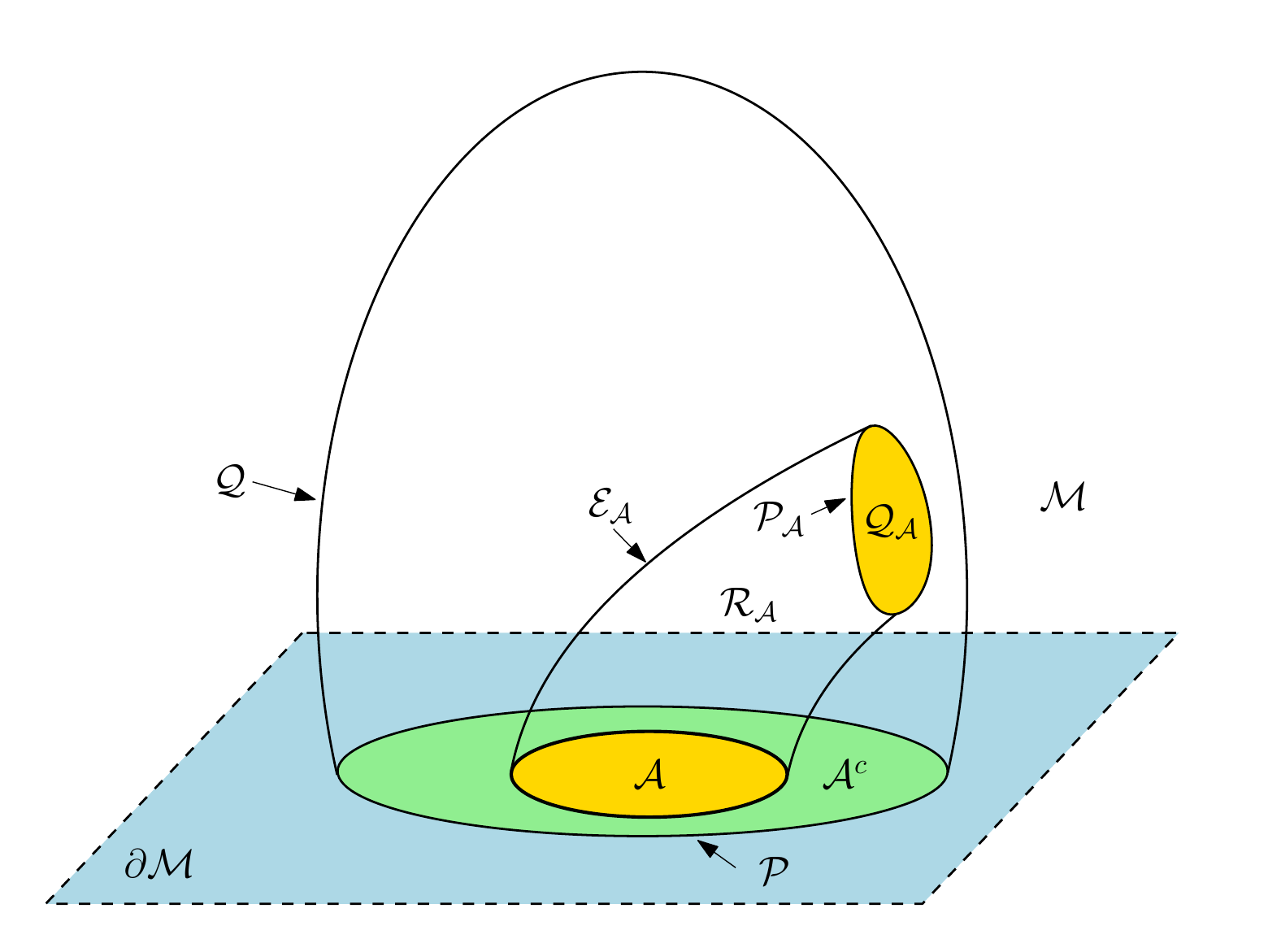} }
\subfloat[AdS black hole]{
\includegraphics[scale=0.5]{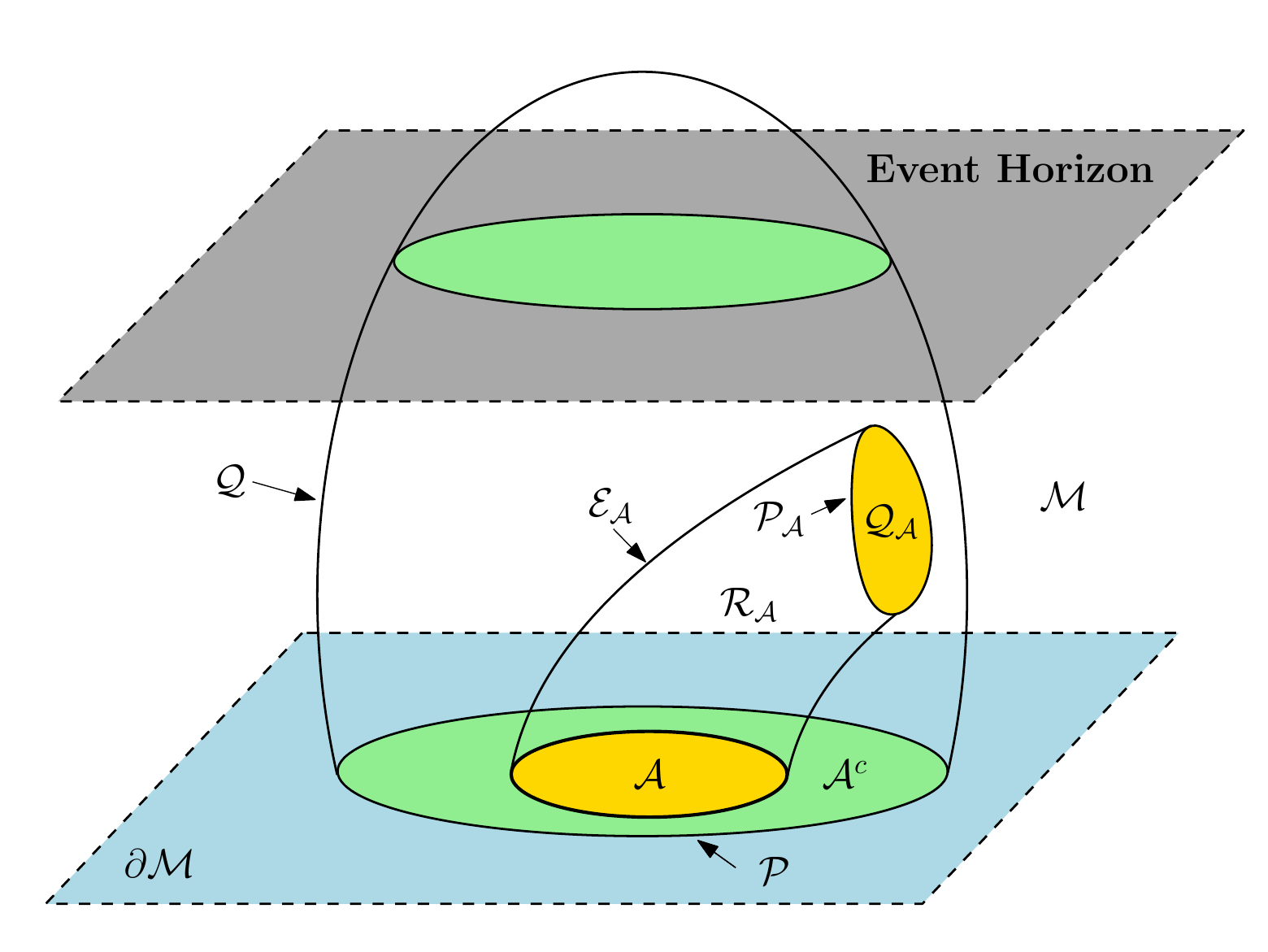} } \caption{Spacetime
setup for the holographic BQFT with an entangled region $\mathcal{A}$. The
minimal surface $\mathcal{E}_{\mathcal{A}}$ could end not only on the
conformal boundary $\partial\mathcal{M}$ but also on the geometric boundary
$\mathcal{Q}$. (a) The bulk manifold is a pure AdS spacetime. The minimal
surface for ${\mathcal{A}}$ and ${\mathcal{A}^{c}}$ have the same homology.
(b) The bulk manifold is an asymptotically AdS black hole. The minimal surface
for ${\mathcal{A}}$ and ${\mathcal{A}^{c}}$ have the different homologies due
to the black hole horizon.}%
\label{Setup2}%
\end{figure}

We consider a spatial region $\mathcal{A}$ with a boundary $\partial
\mathcal{A}$ lying on a Cauchy slice $\Sigma=\mathcal{A}+\mathcal{A}%
^{c}\subset\partial\mathcal{M}$, with $\mathcal{A}^{c}$ the complementary part
of $\mathcal{A}$, as shown in Fig.\ref{Setup2}(a). The HEE is given by the RT
formula \cite{0603001,0605073},
\begin{equation}
S_{\mathcal{A}}=\min_{X}\frac{Area\left(  \mathcal{E}_{\mathcal{A}}\right)
}{4G_{N}^{\left(  d+2\right)  }}\text{,\ }X=\left\{  \mathcal{E}_{\mathcal{A}%
}\Big\vert~\left.  \mathcal{E}_{\mathcal{A}}\right\vert _{\partial\mathcal{M}%
}=\partial\mathcal{A}\text{; }\exists\mathcal{R}_{\mathcal{A}}\subset
\mathcal{M}\text{, }\partial\mathcal{R}_{\mathcal{A}}=\mathcal{E}%
_{\mathcal{A}}\cup\mathcal{A}\right\}  ,
\end{equation}
where $\mathcal{E}_{\mathcal{A}}$ is a codimension-2 minimal surface anchored
on $\partial\mathcal{A}$ in the $(d+2)$-dimensional bulk spacetime
$\mathcal{M}$. The minimal surface $\mathcal{E}_{\mathcal{A}}$ is required to
satisfy a homology constraint: $\mathcal{E}_{\mathcal{A}}$ is smoothly
retractable to the boundary region $\mathcal{A}$. More precisely, there exists
a codimension-1 region $\mathcal{R}_{\mathcal{A}}\subset\mathcal{M}$, the so
called entanglement wedge, which is bounded by the minimal surface
$\mathcal{E}_{\mathcal{A}}$ and the entangled region $\mathcal{A}$ on the
conformal boundary $\partial\mathcal{M}$.

In the presence of the geometric boundary $\mathcal{Q}$, the minimal surface
$\mathcal{E}_{\mathcal{A}}$ could end not only on the conformal boundary
$\partial\mathcal{M}$ but also on the geometric boundary $\mathcal{Q}$ as
showed in \cite{1701.04275,1701.07202}. We thus propose the following formula
for the HEE in BQFT,
\begin{equation}
S_{EE}^{\mathcal{A}}=\min_{X}\frac{Area\left(  \mathcal{E}_{\mathcal{A}%
}\right)  }{4G_{N}^{\left(  d+2\right)  }}\text{, }X=\left\{  \mathcal{E}%
_{\mathcal{A}}\Big\vert~\left.  \mathcal{E}_{\mathcal{A}}\right\vert
_{\partial\mathcal{M}}=\partial\mathcal{A}\text{, }\left.  \mathcal{E}%
_{\mathcal{A}}\right\vert _{Q}=\mathcal{P}_{\mathcal{A}}\text{; }%
\exists\mathcal{R}_{\mathcal{A}}\subset\mathcal{M}\text{, }\partial
\mathcal{R}_{\mathcal{A}}=\mathcal{E}_{\mathcal{A}}\cup\mathcal{A}\cup
Q_{\mathcal{A}}\right\}  ,
\end{equation}
where $\mathcal{P}_{\mathcal{A}}$ divides the geometric boundary $\mathcal{Q}$
into two parts, $\mathcal{Q}_{\mathcal{A}}$ and $\mathcal{Q}_{\mathcal{A}^{c}%
}$, with $\mathcal{Q}_{\mathcal{A}}$ having the same homology with
$\mathcal{A}$ and $\mathcal{Q}_{\mathcal{A}^{c}}$ having the same homology
with $\mathcal{A}^{c}$, as shown in Fig.\ref{Setup2}. Requiring the boundary
condition (\ref{BC}) to be smooth, $\mathcal{E}_{\mathcal{A}}$ should be
orthogonal to $\mathcal{Q}$ when they intersect as showed in
\cite{1701.04275,1701.07202}.

To be concrete, in this work, we consider a bulk spacetime $\mathcal{M}$ with
two boundaries $\mathcal{Q}_{L,R}$ which intersect the conformal boundary
$\partial\mathcal{M}$ at $\mathcal{P}=\pm l/2$ perpendicularly. We choose the
region $\mathcal{A}\subset\partial\mathcal{M}$ as an infinite long strip,
\begin{equation}
x_{1}\in\left[  x-\frac{a}{2},x+\frac{a}{2}\right]  \text{, }x_{i}
\in\mathbb{R}^{d-1}\text{ for }i=2,\cdots d,
\end{equation}
which preserves $(d-1)$-dimensional translation invariance in the directions
$x_{i}$ for $i=2,\cdots d$.

In the static gauge, we can write down the ansatz for the minimal surface
$\mathcal{E}_{\mathcal{A}}$,
\begin{equation}
z=z\left(  x_{1}\right)  \text{, }z\left(  x\pm\frac{a}{2}\right)  =0\text{,
}z\left(  x\right)  =z_{0}\text{, }z^{\prime}\left(  x\right)  =0,
\end{equation}
where $x_{1}=x$ is the turning point of the minimal surface $\mathcal{E}%
_{\mathcal{A}}$.

For a general $(d+2)$-dimensional bulk metric,
\begin{equation}
ds^{2}=-g_{tt}\left(  z\right)  dt^{2}+\sum_{i=1}^{d}g_{ii}\left(  z\right)
dx_{i}^{2}+g_{zz}\left(  z\right)  dz^{2}\text{, }i=1,\cdots d,
\end{equation}
the size $a$ and the HEE $S_{EE}^{\mathcal{A}}$ of the entangled region
$\mathcal{A}$ can be calculated as
\begin{align}
a  &  =2z_{0}\int_{0}^{1}dv\left[  \frac{g_{11}\left(  z_{0}v\right)  }%
{g_{zz}\left(  z_{0}v\right)  }\left(  \frac{\tilde{g}^{2}\left(
z_{0}v\right)  }{\tilde{g}^{2}\left(  z_{0}\right)  }-1\right)  \right]
^{-1/2},\label{a}\\
S_{EE}^{\mathcal{A}}  &  =\frac{l_{AdS}^{d}L^{d-1}}{2G_{N}^{(d+2)}}\int%
_{0}^{1}dv\text{ }z_{0} \tilde{g}\left(  z_{0}v\right)  \left[  \frac
{g_{11}\left(  z_{0}v\right)  }{g_{zz}\left(  z_{0}v\right)  }\left(
1-\frac{\tilde{g}^{2}\left(  z_{0}\right)  }{\tilde{g}^{2}\left(
z_{0}v\right)  }\right)  \right]  ^{-1/2}, \label{SEE}%
\end{align}
where $v=z/z_{0}$ and
\begin{equation}
\tilde{g}\left(  z\right)  =\sqrt{%
{\displaystyle\prod\limits_{i=1}^{d}}
g_{ii}\left(  z\right)  },
\end{equation}
and $L$ is the length of the directions in which the translation invariance is
preserved,
\begin{equation}
\int_{\mathbb{R}^{d-1}}d^{d-1}\mathbf{x}=L^{d-1}.
\end{equation}
Using Eqs.(\ref{a}) and (\ref{SEE}), the HEE can be solved in term of the size
$a$ as $S_{EE}^{\mathcal{A}}(a)$ in principle.

\subsection{Pure AdS Spacetime}

We first consider the bulk spacetime $\mathcal{M}$ as a $(d+2)$-dimensional
pure AdS spacetime with the metric (\ref{AdS}), and choose the geometric
boundary $\mathcal{Q}$ as a $(d+1)$-dimensional hepersurface with the metric
(\ref{AdS-Q}). This is dual to BQFT at the zero temperature.

In the case of pure AdS spacetime, the size $a$ can be integrated to obtain
\begin{equation}
a=2z_{0}\int_{0}^{1}\frac{v^{d}dv}{\sqrt{1-v^{2d}}}=2z_{0}\sqrt{\pi}
\frac{\Gamma\left(  \frac{d+1}{2d}\right)  }{\Gamma\left(  \frac{1}%
{2d}\right)  }.
\end{equation}
The HEE is divergent near the boundary at $v\rightarrow0$. We thus need to
regulate the HEE by putting a small cut-off $\epsilon\ll1$. After the
regulation, the HEE can be obtained as
\begin{equation}
S_{EE}^{\mathcal{A}}=\frac{l_{AdS}^{d}}{2\left(  d-1\right)  G_{N}^{(d+2)}%
}\left[  \left(  \frac{L}{\epsilon}\right)  ^{d-1}-\left(  \frac{L}{z_{0}%
}\right)  ^{d}\frac{a}{2L}\right]  ,
\end{equation}
where the divergent term is proportional to the boundary of the entangled
region $\mathcal{A}$, i.e. $S_{EE}^{\mathcal{A}}\sim L^{d-1}\sim
\partial\mathcal{A}$, as expected. The remaining term is finite.

\begin{figure}[ptb]
\subfloat[Sunset]{
\includegraphics[scale=0.5]{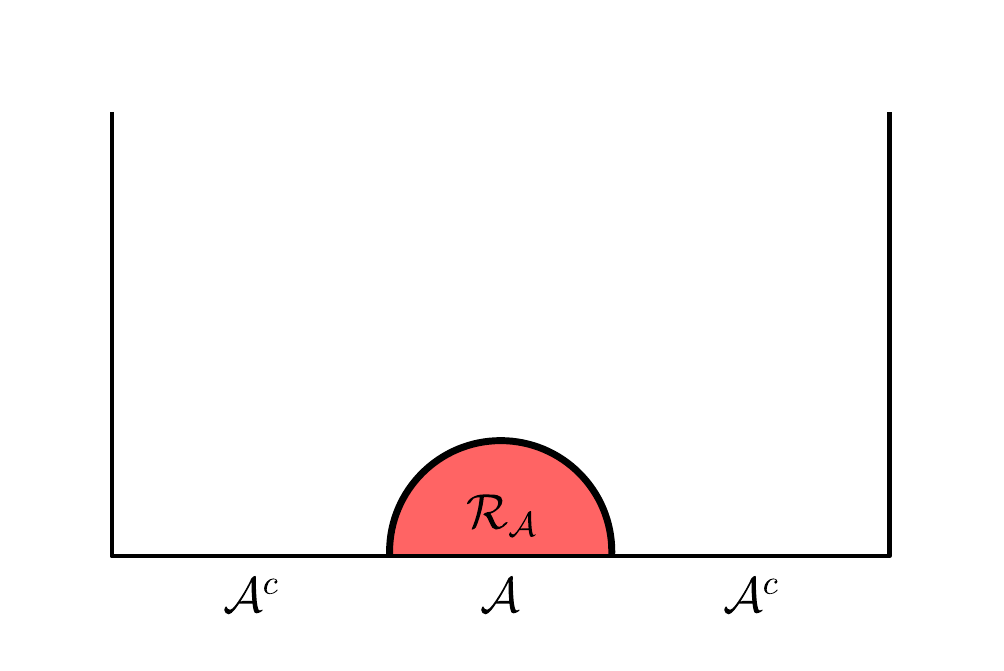} } \subfloat[Sky]{
\includegraphics[scale=0.5]{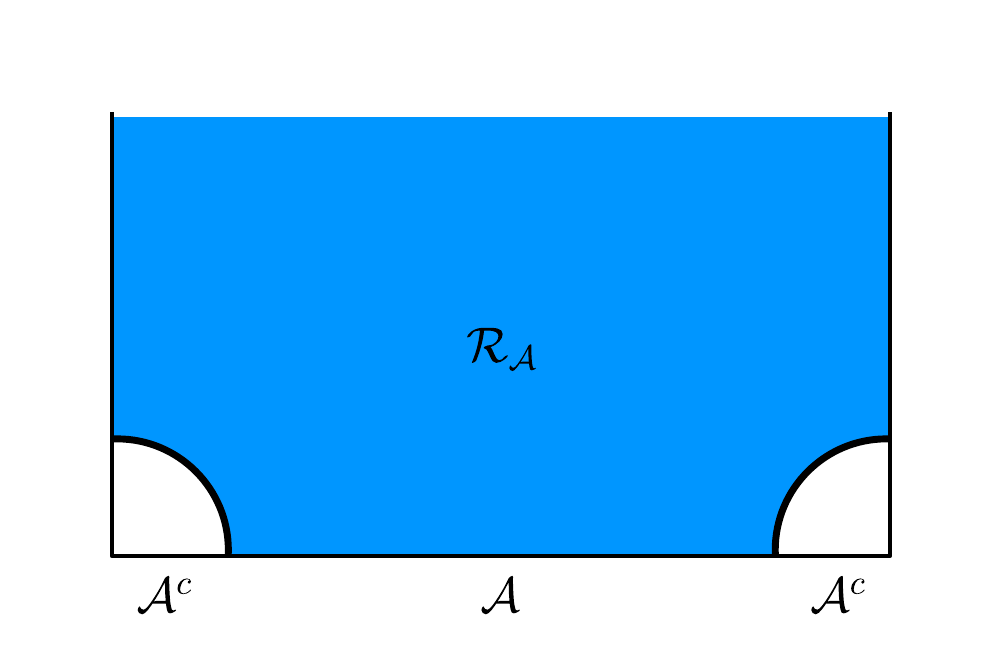} } \subfloat[Rainbow]{
\includegraphics[scale=0.5]{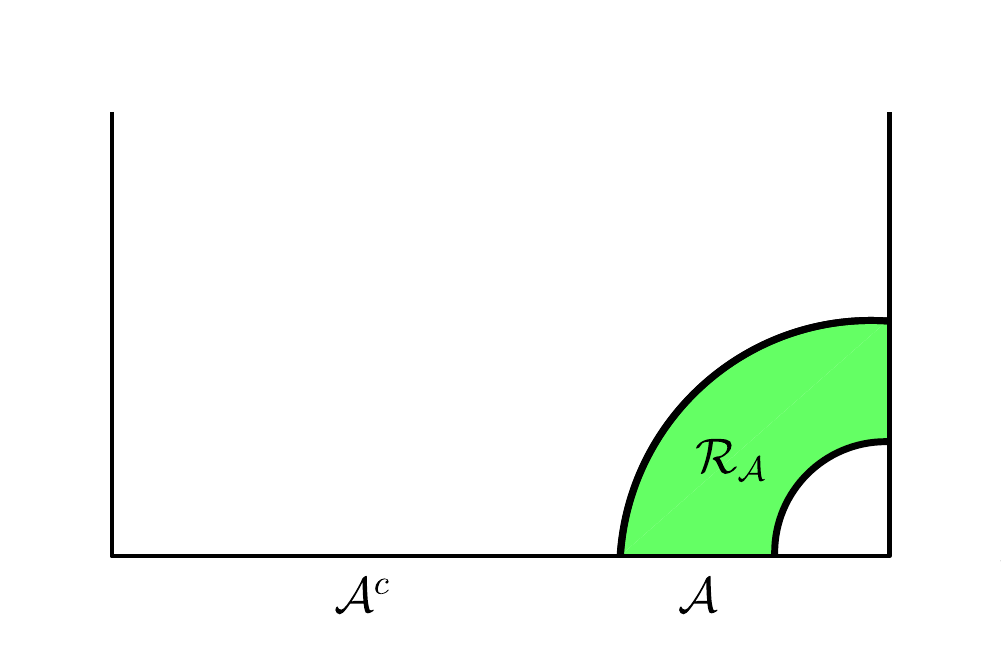} } \caption{The
minimal surfaces in the pure AdS bulk spacetime. (a)  the entanglement wedge
$\mathcal{R}_{\mathcal{A}}$ has the shape of the sunset when the entangled
region $\mathcal{A}$ is very small. (b)  the entanglement wedge $\mathcal{R}%
_{\mathcal{A}}$ has the shape of the sky when the entangled region
$\mathcal{A}$ is very large. (c)  the entanglement wedge $\mathcal{R}%
_{\mathcal{A}}$ has the shape of the rainbow when the entangled region
$\mathcal{A}$ is very closed to the boundary.}%
\label{HEEshapes}%
\end{figure}

In the presence of the geometric boundaries $\mathcal{Q}_{L,R}$, the minimal
surface $\mathcal{E}_{\mathcal{A}}$ could anchor on $\mathcal{Q}_{L,R}$ in
addition to the conformal boundary $\partial\mathcal{M}$. In this work, we
only consider the entangled region $\mathcal{A}$ being a simple connected
region. Under this consideration, there are three types of the minimal
surfaces which satisfy the homology constraint, as shown in
Fig.\ref{HEEshapes}. If the minimal surface $\mathcal{E}_{\mathcal{A}}$ is
connected and only anchors on the conformal boundary $\partial\mathcal{M}$,
the entanglement wedge $\mathcal{R}_{\mathcal{A}}$\ takes the shape of the
sunset as shown in Fig.\ref{HEEshapes}(a). If the minimal surface
$\mathcal{E}_{\mathcal{A}}$ is disconnected and each part anchors on the
different geometric boundaries $\mathcal{Q}_{L,R}$ in addition to the
conformal boundary $\partial\mathcal{M}$, the entanglement wedge
$\mathcal{R}_{\mathcal{A}}$\ takes the shape of the sky as shown in
Fig.\ref{HEEshapes}(b). Finally, if the minimal surface $\mathcal{E}%
_{\mathcal{A}}$ is disconnected and both part anchor on the same geometric
boundary in addition to $\partial\mathcal{M}$, the entanglement wedge
$\mathcal{R}_{\mathcal{A}}$ takes the shape of the rainbow as shown in
Fig.\ref{HEEshapes}(c). The HEE corresponding to the different minimal
surfaces can be calculated as
\begin{align}
S_{sunset}^{\mathcal{A}} &  =S_{EE}^{\mathcal{A}}(a),\\
S_{sky}^{\mathcal{A}} &  =\frac{1}{2}{S_{EE}^{\mathcal{A}}(l-a+2|x|)+\frac
{1}{2}S_{EE}^{\mathcal{A}}(l-a-2|x|)},\\
S_{rainbow}^{\mathcal{A}} &  =\frac{1}{2}{S_{EE}^{\mathcal{A}}(l+a-2|x|)+\frac
{1}{2}S_{EE}^{\mathcal{A}}(l-a-2|x|)}.
\end{align}
Although each of the HEE in the above three cases is the local minimum, the
global minimum depends on the size $a$ and the location $x$ of the entangled
region $\mathcal{A}$.

For a small enough $\mathcal{A}$, the entanglement wedge $\mathcal{R}%
_{\mathcal{A}}$\ takes the shape of the sunset. While for a large enough
$\mathcal{A}$, the entanglement wedge $\mathcal{R}_{\mathcal{A}}$\ takes the
shape of the sky. If the location of the region $\mathcal{A}$ is very close to
one of the geometric boundaries $\mathcal{Q}_{L,R}$, the minimal surface
$\mathcal{E}_{\mathcal{A}}$ would be inclined to the boundary and break into
two parts, the entanglement wedge $\mathcal{R}_{\mathcal{A}}$ will take the
shape of the rainbow as shown in Fig.\ref{HEEshapes}(c).

\begin{figure}[ptb]
\centerline{\includegraphics[scale=0.5]{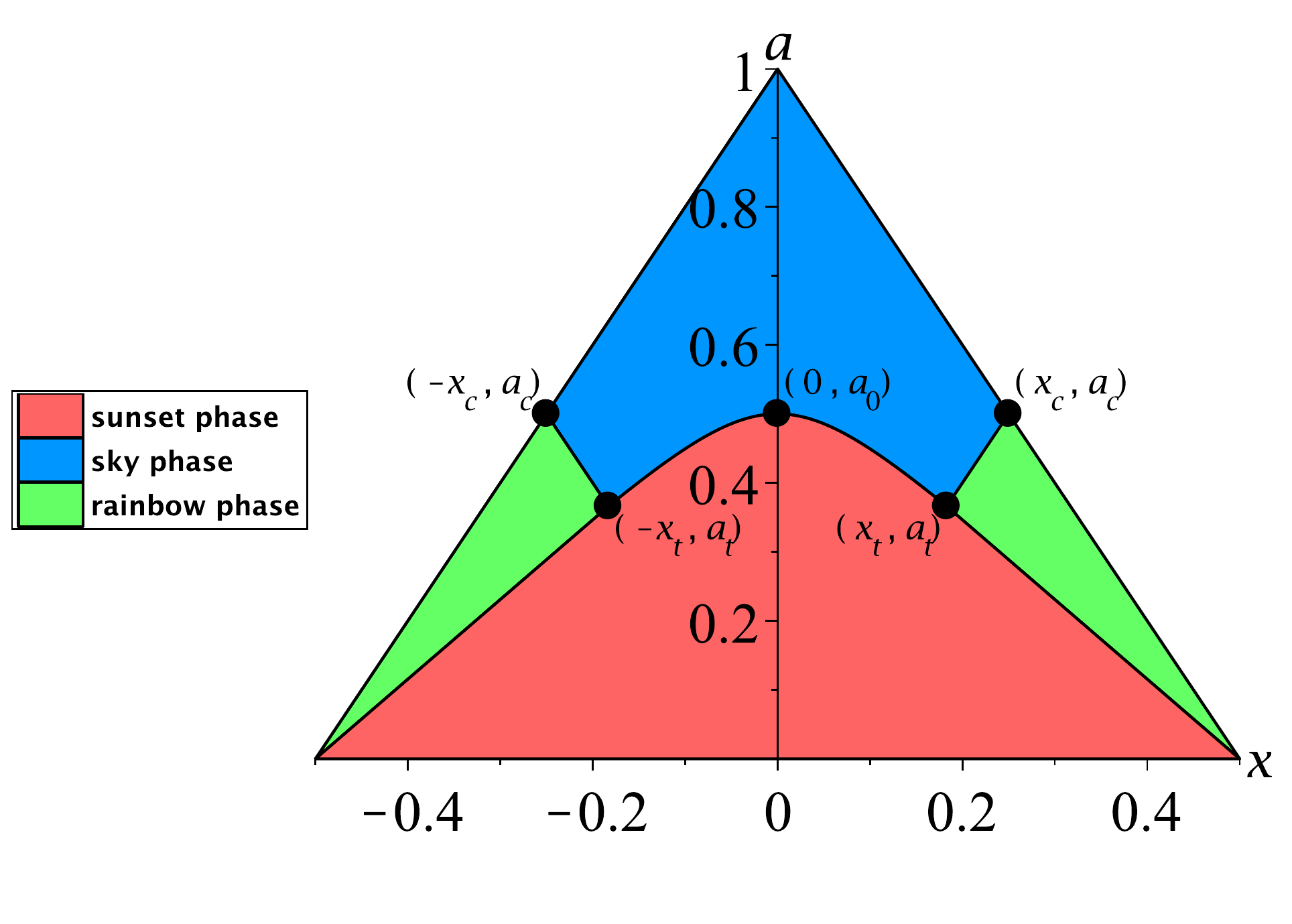}} \caption{Phase
Diagram of the holographic entanglement entropy in the pure AdS bulk
spacetime. Different phases are marked with the different colors.}%
\label{PureAdsPD}%
\end{figure}

The HEE transfers among the three phases as the size $a$ and the location $x$
of the entangled region $\mathcal{A}$ varying, which correspond to the quantum
phase transitions at the zero temperature in the dual BQFT. The phase diagram
is shown in Fig.\ref{PureAdsPD} (We set $l=1/2$ in the figures of this paper).
At the middle, $x=0$, there is a critical value $a_{0}$ for the size of the
entangled region $\mathcal{A}$. For $a<a_{0}$, the entanglement wedge
$\mathcal{R}_{\mathcal{A}}$ takes the shape of the sunset, while for $a>a_{0}%
$, the minimal surface $\mathcal{E}_{\mathcal{A}}$ breaks into two parts and
the entanglement wedge $\mathcal{R}_{\mathcal{A}}$ takes the shape of the sky.
When $x$ is away from the middle, the critical value decreases until it
reaches the triple critical points at $\left(  \pm x_{t},a_{t}\right)  $ where
a new phase, in which the entanglement wedge $\mathcal{R}_{\mathcal{A}}$ takes
the shape of the rainbow, emerges due to the effect of the boundary
$\mathcal{Q}$. When $\left\vert x\right\vert $ is beyond the critical points
at $\left(  \pm x_{c},a_{c}\right)  $, the sky phase disappears, the sunset
phase and the rainbow phase compete until $x$ reaches the boundaries at $x=\pm
l/2$.

In the pure AdS case, it is easy to see that, for a entangled region
$\mathcal{A}$, and its complementary $\mathcal{A}^{c}$, the associated minimal
surfaces $\mathcal{E}_{\mathcal{A}}$ and $\mathcal{E}_{\mathcal{A}^{c}}$ have
the same homology. Therefore, $\mathcal{A}$ and $\mathcal{A}^{c}$ share the
same minimal surface $\mathcal{E}_{\mathcal{A}}=\mathcal{E}_{\mathcal{A}^{c}}%
$, as well as the same HEE $S_{EE}^{\mathcal{A}} = S_{EE}^{\mathcal{A}^{c}}$.

\subsection{Schwarzschild-AdS Black Hole}

We next consider the bulk spacetime $\mathcal{M}$ as a $(d+2)$-dimensional
Schwarzschild-AdS spacetime with the metric (\ref{AdS-Schw}), and choose the
geometric boundary $\mathcal{Q}$ as a $(d+1)$-dimensional hypersurface embeded
in $\mathcal{M}$ with the metric (\ref{AdS-Schw-Q}). This is dual to BQFT at
the finite temperature. The temperature in BQFT is identified with the Hawking
temperature of the black hole by the holographic correspondence. The
temperature and the entropy density of the black hole were given in Eq.
(\ref{T and S}).

In the case of AdS black hole spacetime, the size $a$ of the entangled region
$\mathcal{A}$ can be expressed as the following integral
\begin{equation}
a=2z_{0}\int_{0}^{1}\frac{v^{d}dv}{\sqrt{\left(  1-\left(  bv\right)
^{d+1}\right)  \left(  1-v^{2d}\right)  }},
\end{equation}
where we have defined the parameter $b=z_{0}/z_{H}$, which measures how close
the minimal surface $\mathcal{E}_{\mathcal{A}}$ is from the horizon.

Similar to the pure AdS case, the HEE is divergent near the boundary at $v
\rightarrow0$ and we need to regulate it by putting a small cut-off $\epsilon
$. After the regulation, the HEE can be obtained as
\begin{align}
S_{EE}  &  =\frac{l_{AdS}^{d}}{2\left(  d-1\right)  G_{N}^{(d+2)}}\left(
\frac{L}{\epsilon}\right)  ^{d-1}-\frac{1}{d-1} \left[  \rule{0cm}{0.95cm}%
\right.  \frac{a}{b^{d}} \frac{S_{BH}}{L}\nonumber\\
&  -b z_{0} \int_{0}^{1}\frac{\left(  d-3\right)  v}{\sqrt{\left(  1-\left(
bv\right)  ^{d+1}\right)  \left(  1-v^{2d}\right)  }}dv-b z_{0} \int_{0}%
^{1}\frac{\left(  d+3\right)  v^{2d+1}}{\sqrt{\left(  1-\left(  bv\right)
^{d+1}\right)  \left(  1-v^{2d}\right)  }}dv\left.  \rule{0cm}{0.95cm}\right]
,
\end{align}
where the first term is divergent and is proportional to the boundary of the
region $\mathcal{A}$ as the same as in the pure AdS case. The remaining terms
in the square brackets are finite.

In the small size limit $a\rightarrow0$, the turning point $z_{0}\rightarrow0$
as well, so that the the parameter $b=z_{0}/z_{H}\rightarrow0$. The HEE thus
reduces to%
\begin{align}
S_{EE}  &  \simeq\frac{l_{AdS}^{d}}{2\left(  d-1\right)  G_{N}^{(d+2)}}\left(
\frac{L}{\epsilon}\right)  ^{d-1}- \frac{a}{\left(  d-1\right)  b^{d}}%
\frac{S_{BH}}{L}\nonumber\\
&  =\frac{l_{AdS}^{d}}{2\left(  d-1\right)  G_{N}^{(d+2)}}\left[  \left(
\frac{L}{\epsilon}\right)  ^{d-1}-\left(  \frac{L}{z_{0}}\right)  ^{d}\frac
{a}{2L}\right]  ,
\end{align}
which is exact the same as in the pure AdS case as it should be. While for a
finite size $a$, the HEE in the black hole case dramatically deviates from
that in the pure AdS case.

\begin{figure}[ptb]
\subfloat[Sunset]{
\includegraphics[scale=0.5]{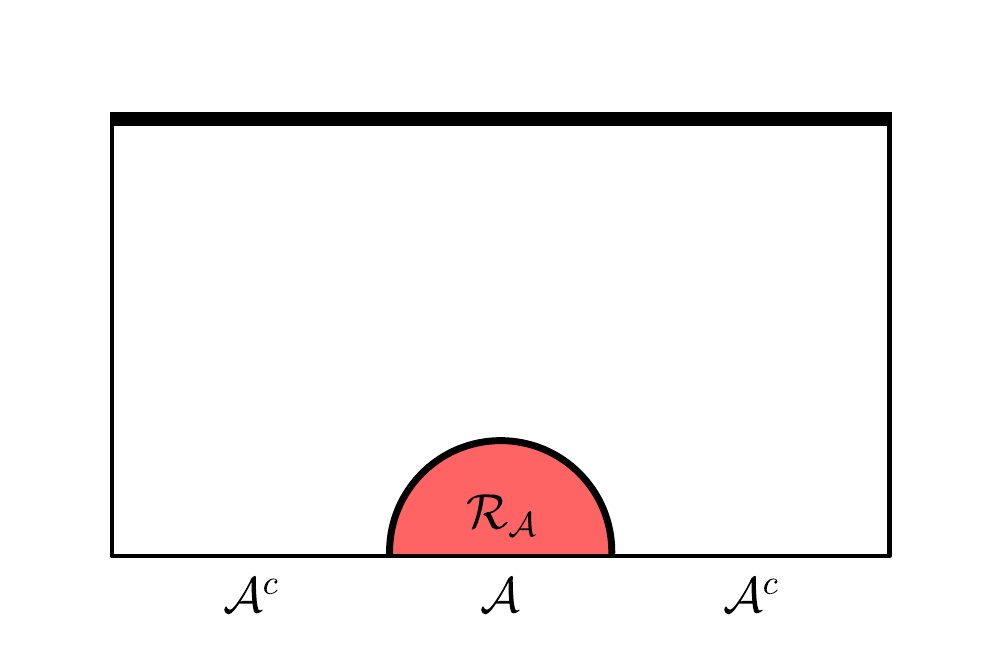} } \subfloat[Sky]{
\includegraphics[scale=0.5]{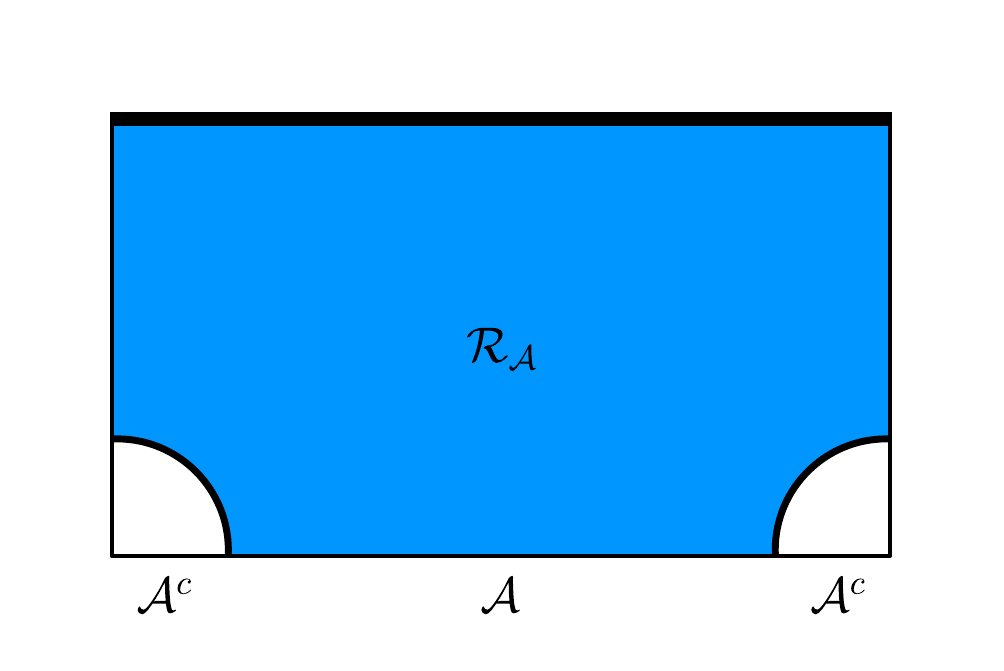} } \subfloat[Rainbow]{
\includegraphics[scale=0.5]{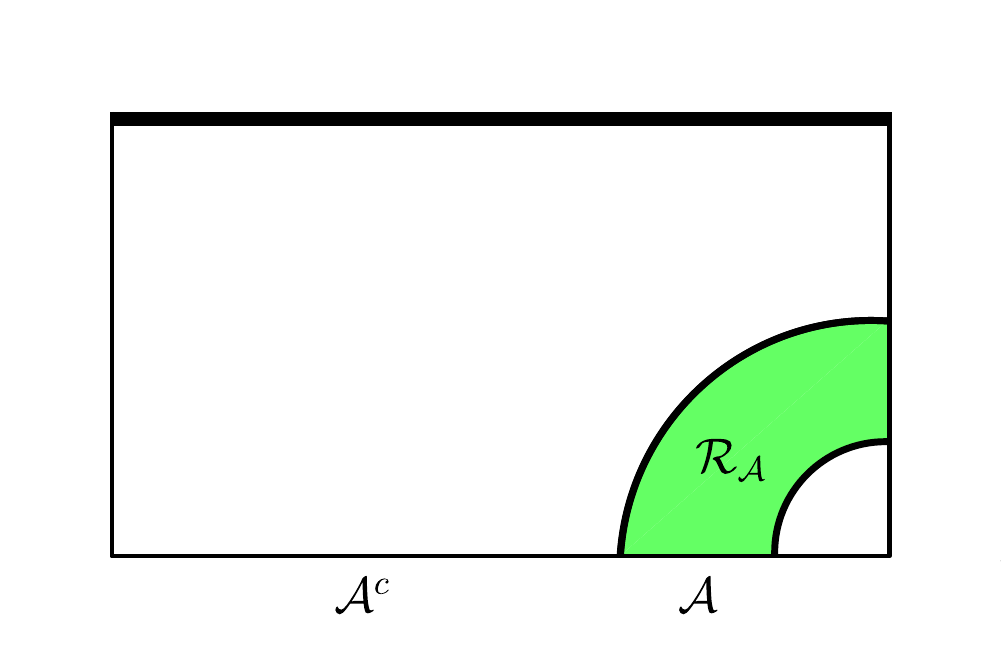} } \hfill
\subfloat[Reversed-sunset]{
\includegraphics[scale=0.5]{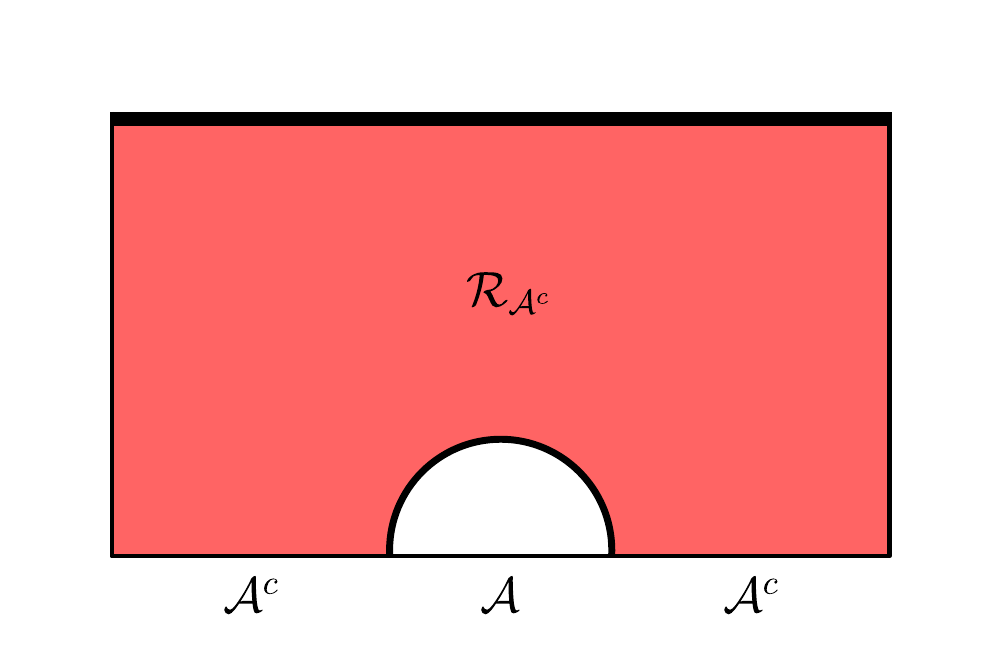} }
\subfloat[Reversed-sky]{
\includegraphics[scale=0.5]{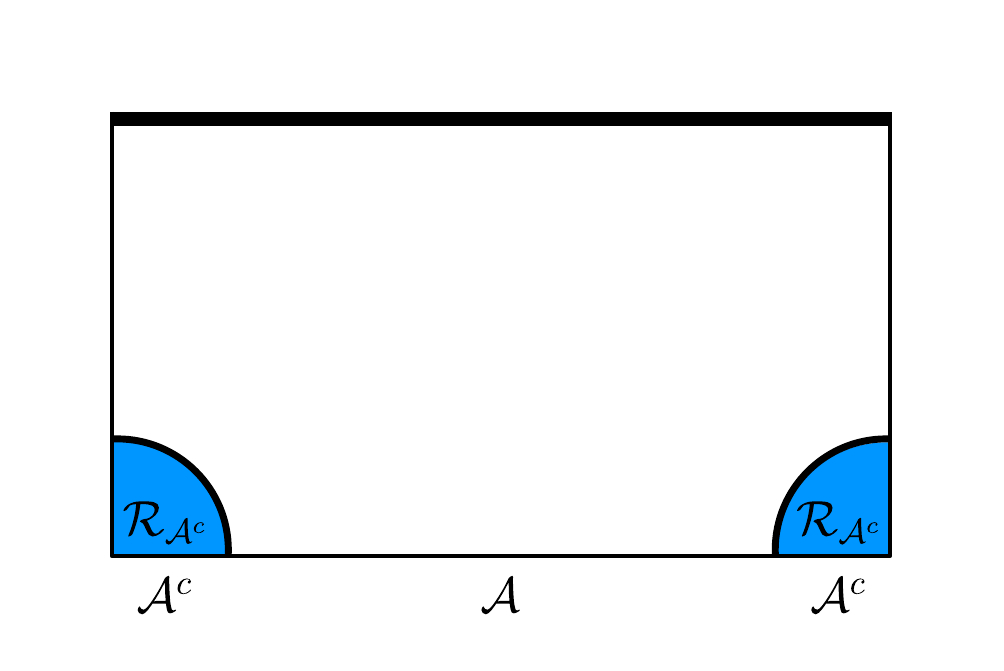} }
\subfloat[Reversed-rainbow]{
\includegraphics[scale=0.5]{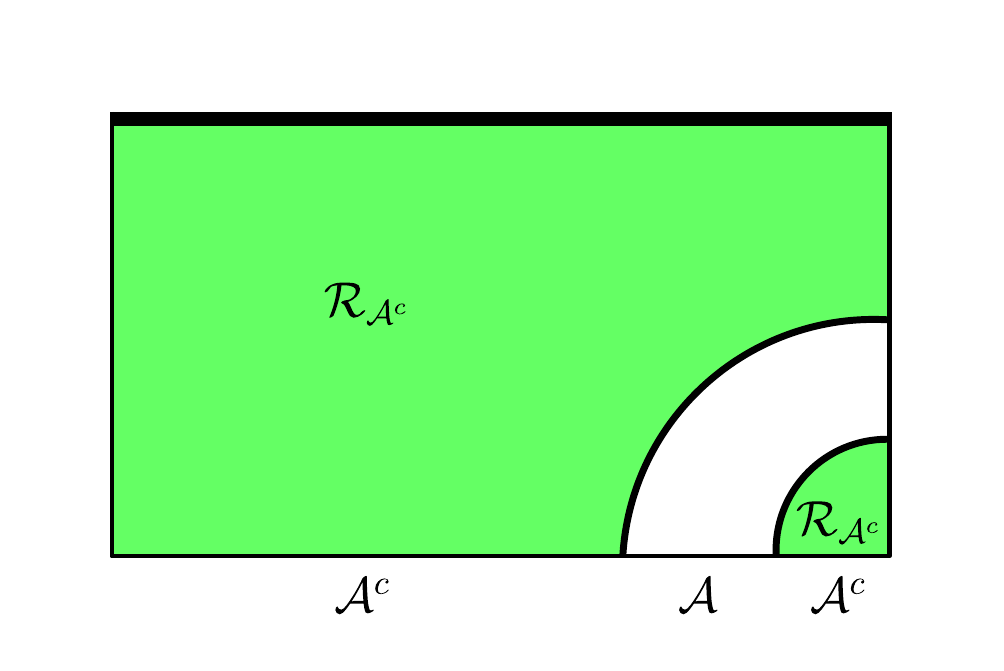} } \caption{The
minimal surfaces in the asymptotically AdS black hole bulk spacetime. The
thick black line at the top indicates the black hole horizon. (a/d)  the
entanglement wedge $\mathcal{R}_{\mathcal{A}}$/$\mathcal{R}_{\mathcal{A}^{c}}$
has the shape of the sunset/reversed-sunset when the entangled region
$\mathcal{A}$ is very small. (b/e)  the entanglement wedge $\mathcal{R}%
_{\mathcal{A}}$/$\mathcal{R}_{\mathcal{A}^{c}}$ has the shape of the
sky/reversed-sky when the entangled region $\mathcal{A}$ is very large. (c/f)
the entanglement wedge $\mathcal{R}_{\mathcal{A}}$/$\mathcal{R}_{\mathcal{A}%
^{c}}$ has the shape of the rainbow/reversed-rainbow when the entangled region
$\mathcal{A}$ is very closed to the boundary.}%
\label{HEEshapesBH}%
\end{figure}

In the case of the black hole spacetime, the associated minimal surfaces
$\mathcal{E}_{\mathcal{A}}$ and $\mathcal{E}_{\mathcal{A}^{c}}$ for the
entangled region $\mathcal{A}$ and its complementary $\mathcal{A}^{c}$ have
different homology due to the presence of the black hole horizon, so that the
HEE $S_{EE}^{\mathcal{A}}$ for a region $\mathcal{A}$ is generically not the
same as the HEE $S_{EE}^{\mathcal{A}^{c}}$ for its complementary
$\mathcal{A}^{c}$. This is the crucial difference between the cases of the AdS
black hole and the pure AdS spacetime.

As in the pure AdS case, there are three types of the minimal surfaces
depending on the size $a$ and the location $x$ of the region $\mathcal{A}$,
and similarly for its complementary $\mathcal{A}^{c}$, as shown in
Fig.\ref{HEEshapesBH}. The HEE corresponding to the different minimal surfaces
$\mathcal{E}_{\mathcal{A}}$ can be calculated as
\begin{align}
S^{\mathcal{A}}_{sunset}  &  = S_{EE}^{\mathcal{A}}(a),\\
S^{\mathcal{A}}_{sky}  &  = \frac{1}{2}{S_{EE}^{\mathcal{A}}(l-a+2|x|)+\frac
{1}{2}S_{EE}^{\mathcal{A}}(l-a-2|x|)} + S_{BH},\\
S^{\mathcal{A}}_{rainbow}  &  = \frac{1}{2}{S_{EE}^{\mathcal{A}}%
(l+a-2|x|)+\frac{1}{2}S_{EE}^{\mathcal{A}}(l-a-2|x|)}.
\end{align}
The HEE corresponding to the minimal surfaces $\mathcal{E}_{\mathcal{A}^{c}}$
can be calculated as
\begin{align}
S^{\mathcal{A}^{c}}_{R-sunset}  &  = S_{EE}^{\mathcal{A}}(a) + S_{BH},\\
S^{\mathcal{A}^{c}}_{R-sky}  &  = \frac{1}{2}{S_{EE}^{\mathcal{A}%
}(l-a+2|x|)+\frac{1}{2}S_{EE}^{\mathcal{A}}(l-a-2|x|)},\\
S^{\mathcal{A}^{c}}_{R-rainbow}  &  = \frac{1}{2}{S_{EE}^{\mathcal{A}%
}(l+a-2|x|)+\frac{1}{2}S_{EE}^{\mathcal{A}}(l-a-2|x|)} + S_{BH}.
\end{align}

For a small enough $\mathcal{A}$, hence the large enough $\mathcal{A}^{c}$,
the minimal surface $\mathcal{E}_{\mathcal{A}}$ only anchors on $\partial
\mathcal{M}$, and the entanglement wedge $\mathcal{R}_{\mathcal{A}}$ takes the
shape of the sunset, similar to the case of the pure AdS; while the minimal
surface $\mathcal{E}_{\mathcal{A}^{c}}$ includes both $\mathcal{E}%
_{\mathcal{A}}$ and the black hole horizon, and the entanglement wedge
$\mathcal{R}_{\mathcal{A}^{c}}$ takes the shape of the reversed-sunset, as
shown in Fig.\ref{HEEshapesBH}(a,d). For a large enough $\mathcal{A}$, hence
the small enough $\mathcal{A}^{c}$, the minimal surface $\mathcal{E}%
_{\mathcal{A}}$ breaks into two parts plus the horizon and  the entanglement
wedge $\mathcal{R}_{\mathcal{A}}$ takes the shape of the sky; while the
minimal surface $\mathcal{E}_{\mathcal{A}^{c}}$ is $\mathcal{E}_{\mathcal{A}}$
minus the horizon, and  the entanglement wedge $\mathcal{R}_{\mathcal{A}^{c}}$
takes the shape of the reversed-sky, as shown in Fig.\ref{HEEshapesBH}(b,e).
When the location of the entangled region $\mathcal{A}$ is very close to one
of the geometric boundaries, the minimal surface $\mathcal{E}_{\mathcal{A}}$
would be inclined to the boundary and  the entanglement wedge $\mathcal{R}%
_{\mathcal{A}}$ takes the shape of the rainbow; while the minimal surface
$\mathcal{E}_{\mathcal{A}^{c}}$ is $\mathcal{E}_{\mathcal{A}}$ plus the
horizon and the entanglement wedge $\mathcal{R}_{\mathcal{A}^{c}}$ takes the
shape of the reversed-rainbow as shown in Fig.\ref{HEEshapesBH}(c,f). However,
the positions of the phase transitions among the three phases are usually
different for $\mathcal{A}$ and $\mathcal{A}^{c}$, that makes the full phase
structure for $\mathcal{A}$ and $\mathcal{A}^{c}$ rather complicated.

The HEE transfers among the different phases corresponding to the phase
transitions at the finite temperature in the dual BQFT. These phase
transitions are the mixture of the quantum phase transition and the thermal
phase transition.

\begin{figure}[ptb]
\centerline{\includegraphics[scale=0.5]{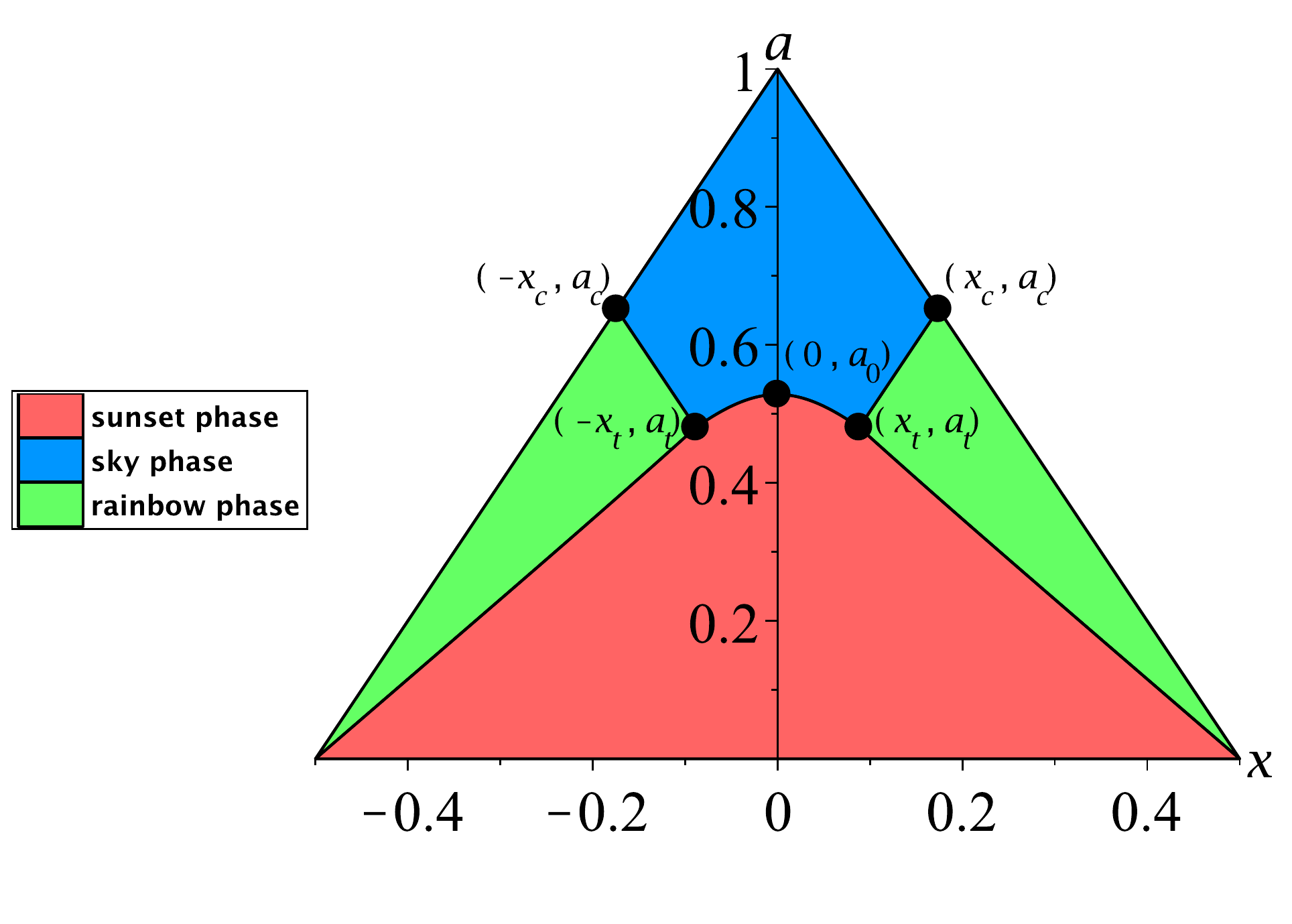}} \caption{Phase
Diagram of the holographic entanglement entropy $S_{EE}^{\mathcal{A}}$ for a
entangled region $\mathcal{A}$ in the Schwarzschild-AdS blackhole bulk
spacetime with the horizon $zH= 1.5$.}%
\label{AdsZh=1p5PD}%
\end{figure}

Fig.\ref{AdsZh=1p5PD} shows the phase diagram of the HEE for the entangled
region\ $\mathcal{A}$ at the horizon $z_{H}=1.5$, i.e. at the temperature
$T=0.212$ in the dual BQFT. At the middle, $x=0$, there is a critical value
$a_{0}$ for the size of the region $\mathcal{A}$. For $a<a_{0}$, the
entanglement wedge $\mathcal{R}_{\mathcal{A}}$ takes the shape of the sunset,
while for $a>a_{0}$, $\mathcal{E}_{\mathcal{A}}$ breaks into two parts plus
the horizon and the entanglement wedge $\mathcal{R}_{\mathcal{A}}$ takes the
shape of the sky. When $x$ is away from the center at $x=0$, the critical
value decreases until it reaches the triple critical points at $\left(  \pm
x_{t},a_{t}\right)  $ where a new phase, in which the entanglement wedge
$\mathcal{R}_{\mathcal{A}}$ takes the shape of the rainbow, emerges due to the
effect of the boundary $\mathcal{Q}$. When $\left\vert x\right\vert $ is
beyond the critical points at $\left(  \pm x_{c},a_{c}\right)  $, the sky
phase disappears, the sunset phase and the rainbow phase compete until $x$
reaches the boundaries at $x=\pm l/2$.

\begin{figure}[ptb]
\subfloat[$z_{H} = 0.6$]{
\includegraphics[scale=0.2]{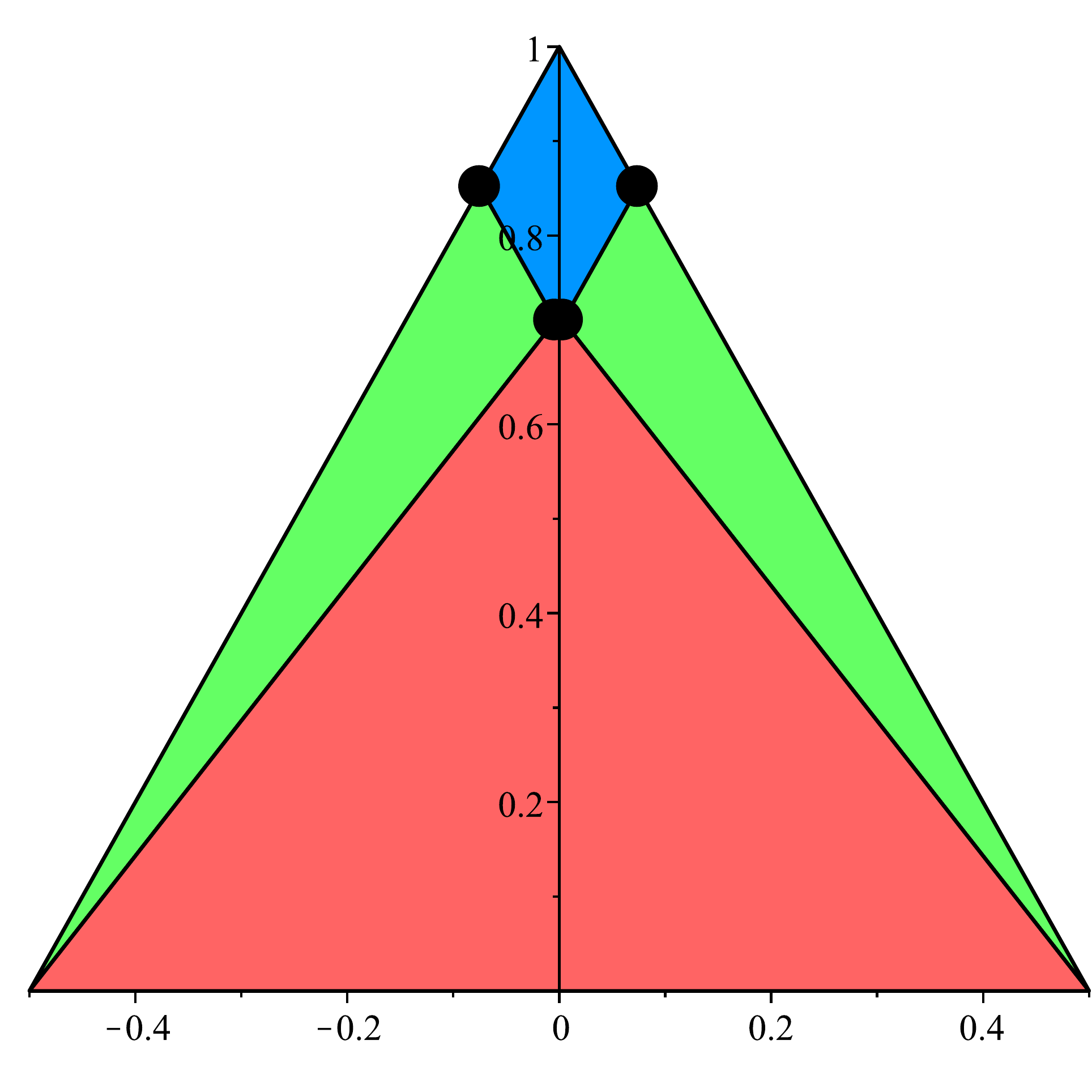} }
\subfloat[$z_{H} = 1$]{
\includegraphics[scale=0.2]{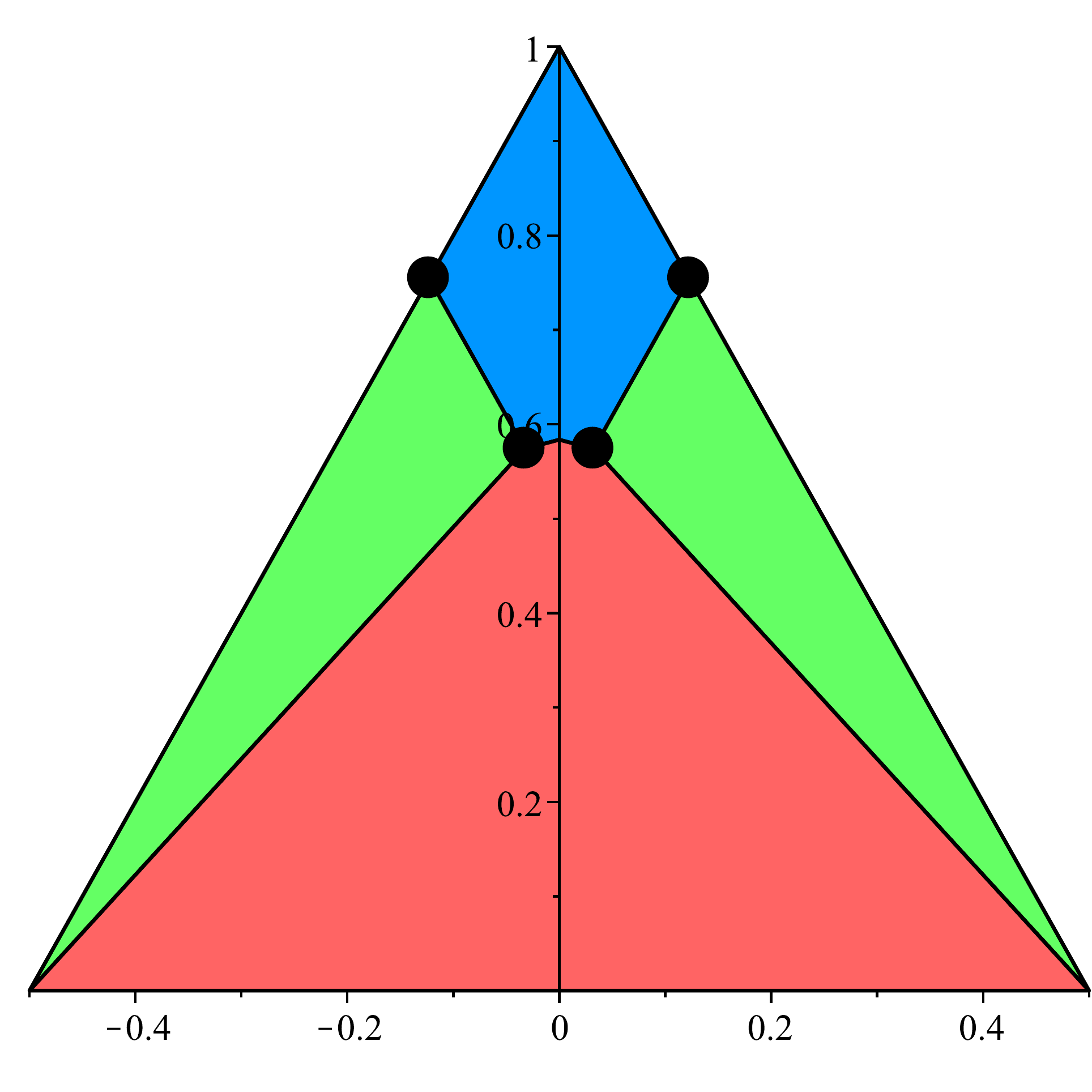} }
\subfloat[$z_{H} = 2$]{
\includegraphics[scale=0.2]{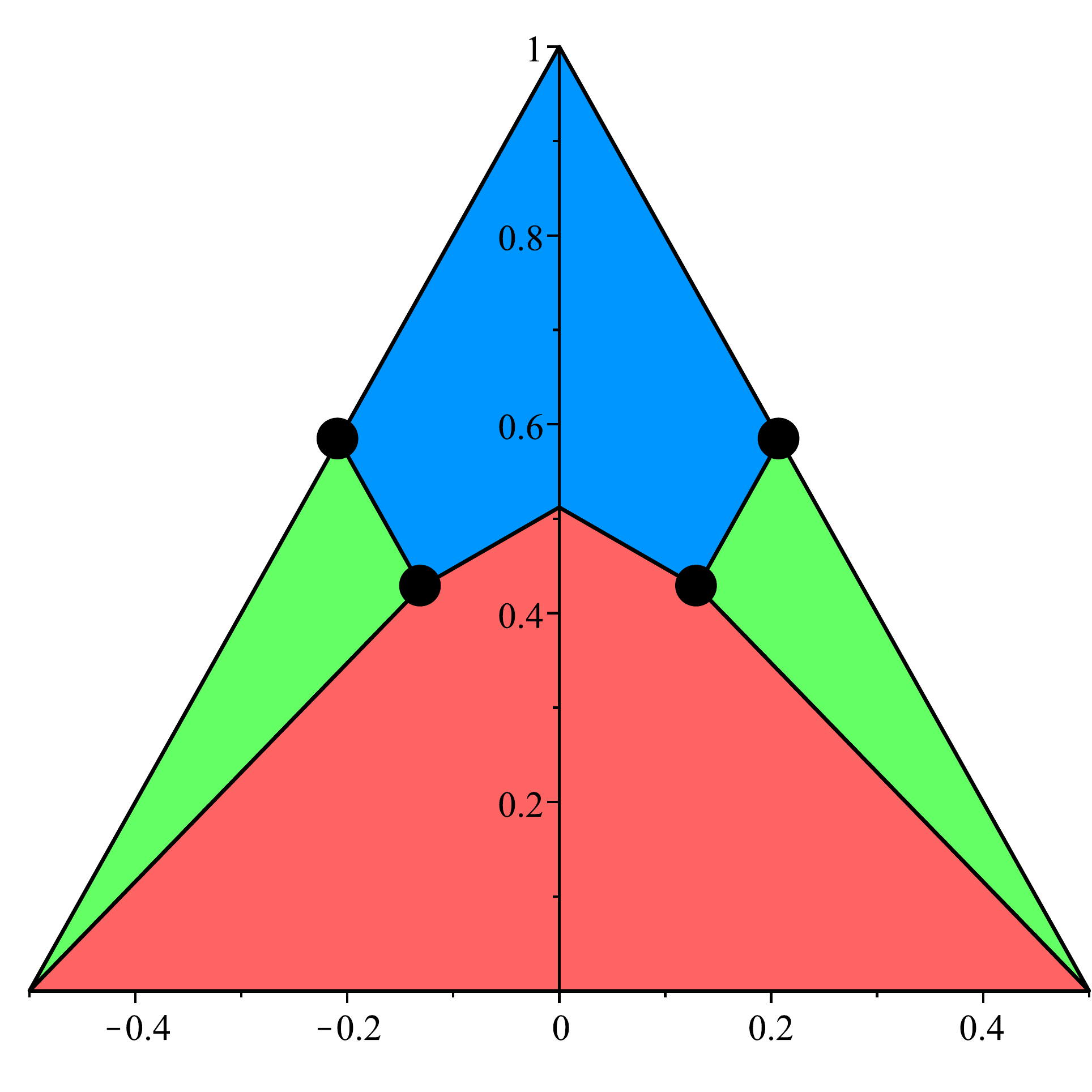} }
\subfloat[$z_{H} = 9$]{
\includegraphics[scale=0.2]{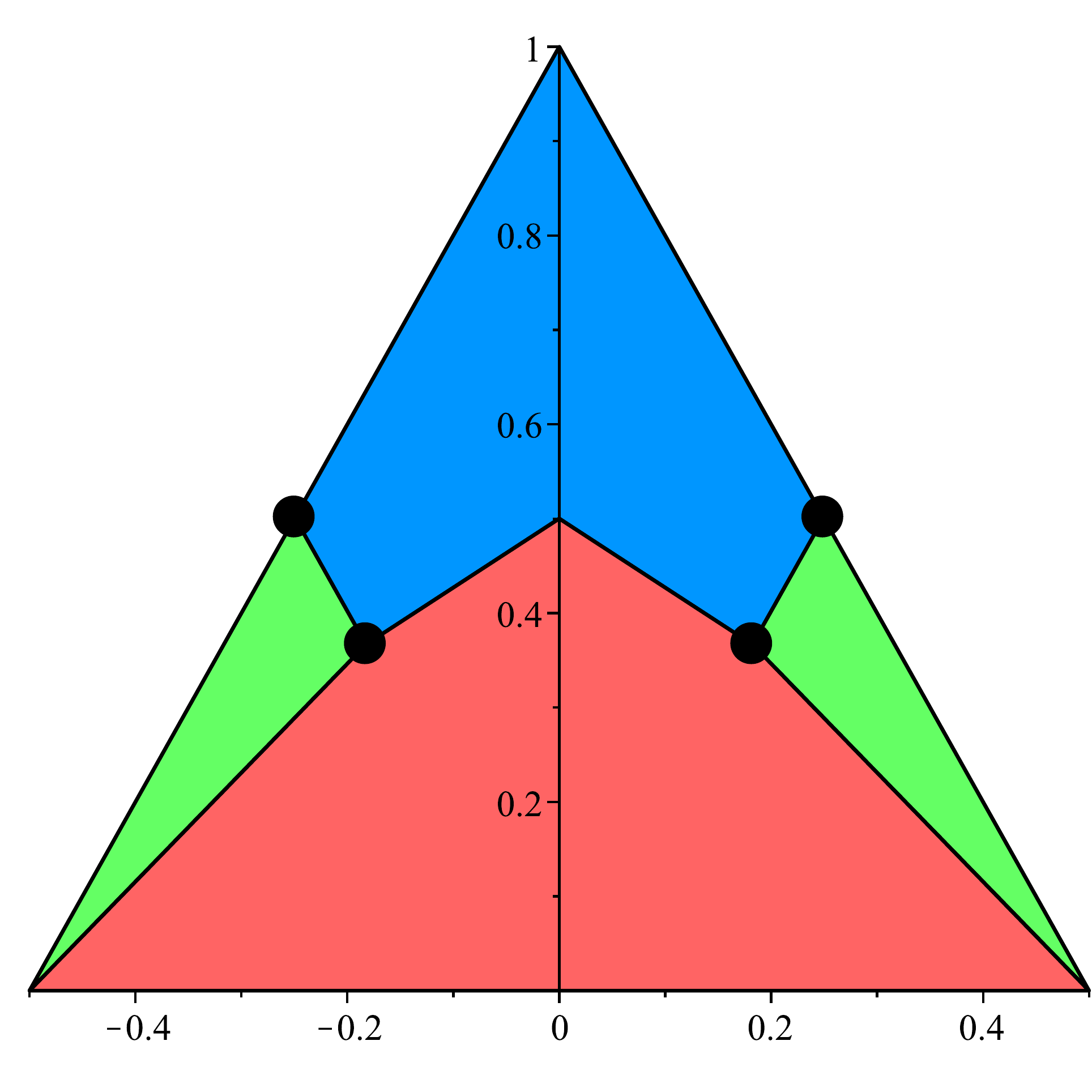} } \hfill
\subfloat[$z_{H} = 0.6$]{
\includegraphics[scale=0.2]{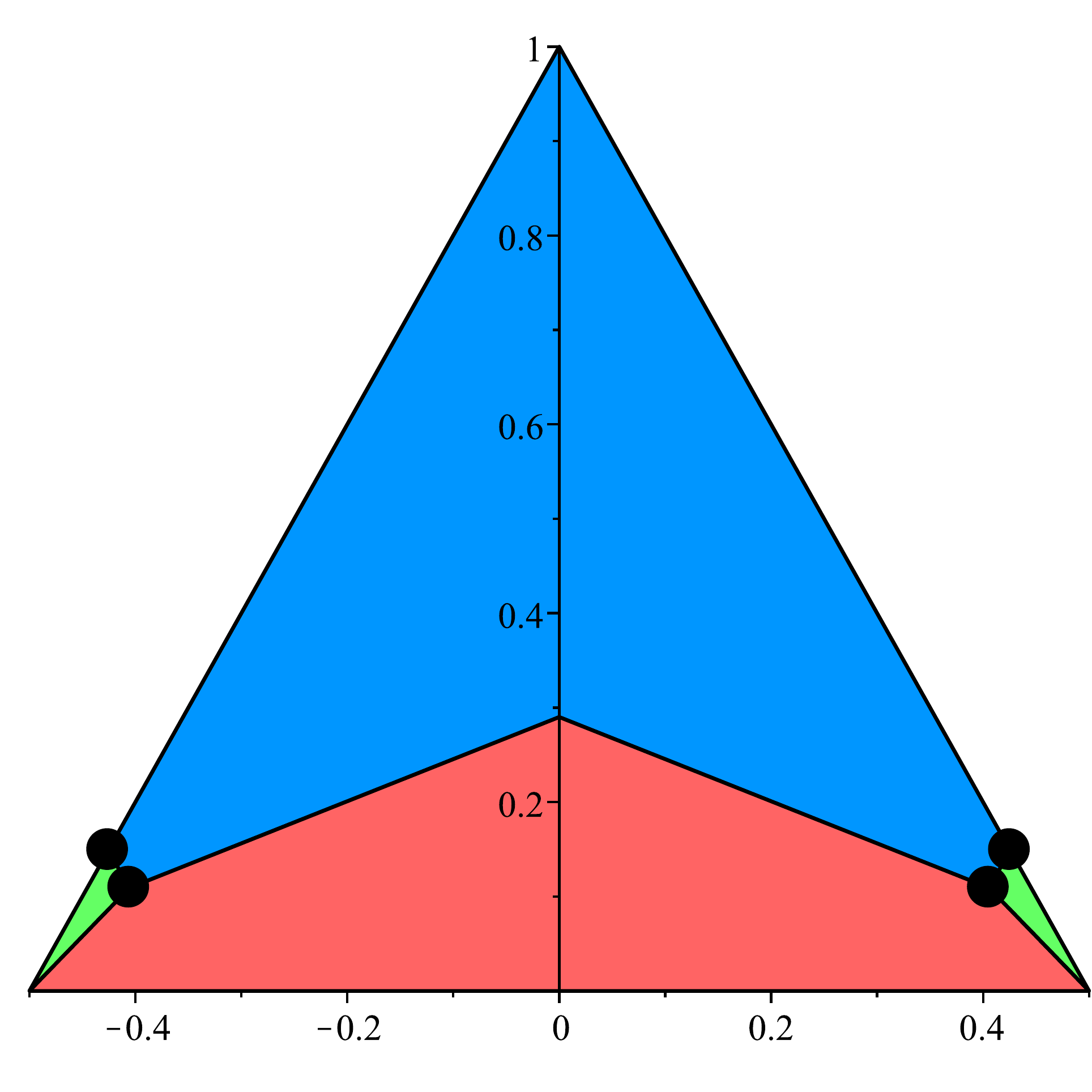} }
\subfloat[$z_{H} = 1$]{
\includegraphics[scale=0.2]{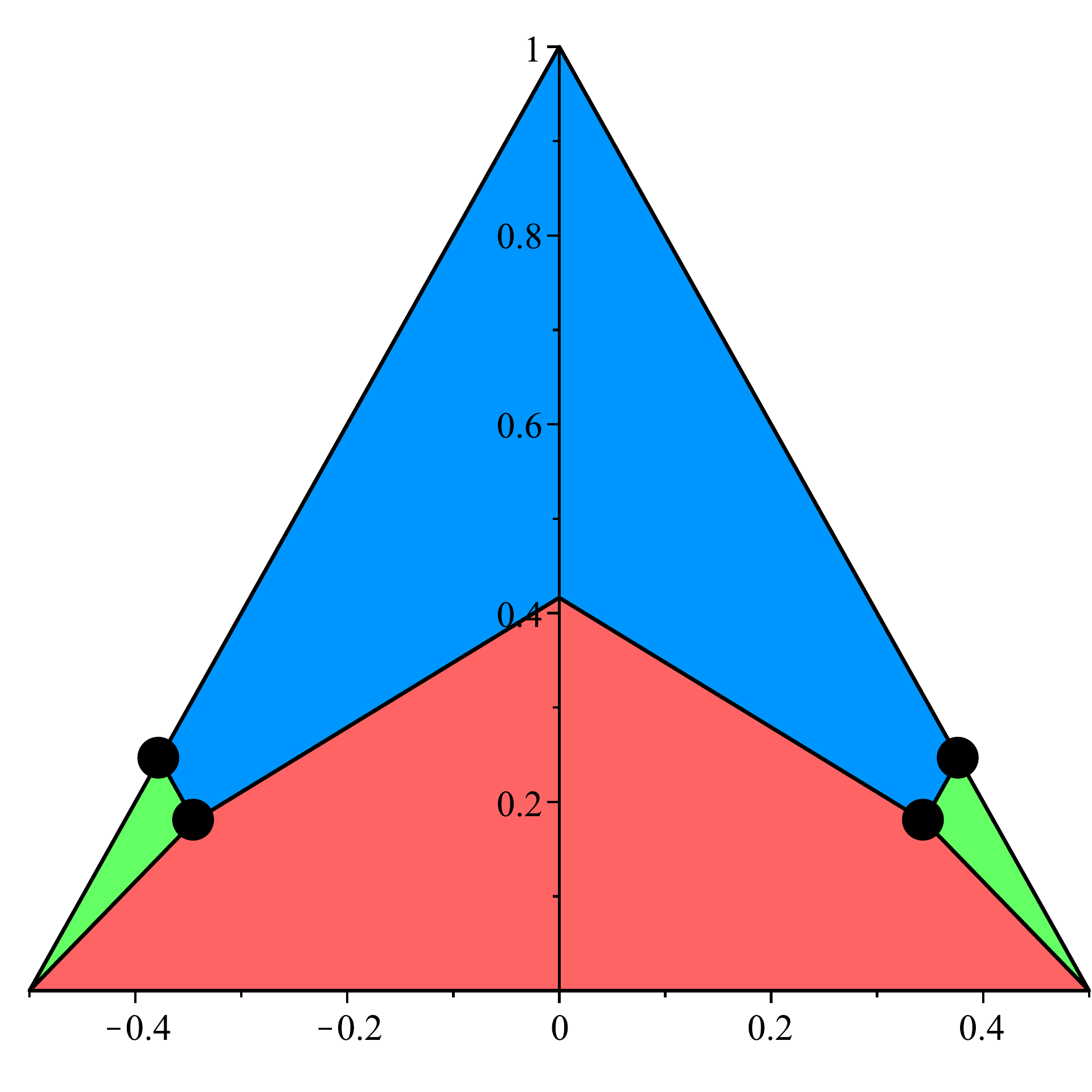} }
\subfloat[$z_{H} = 2$]{
\includegraphics[scale=0.2]{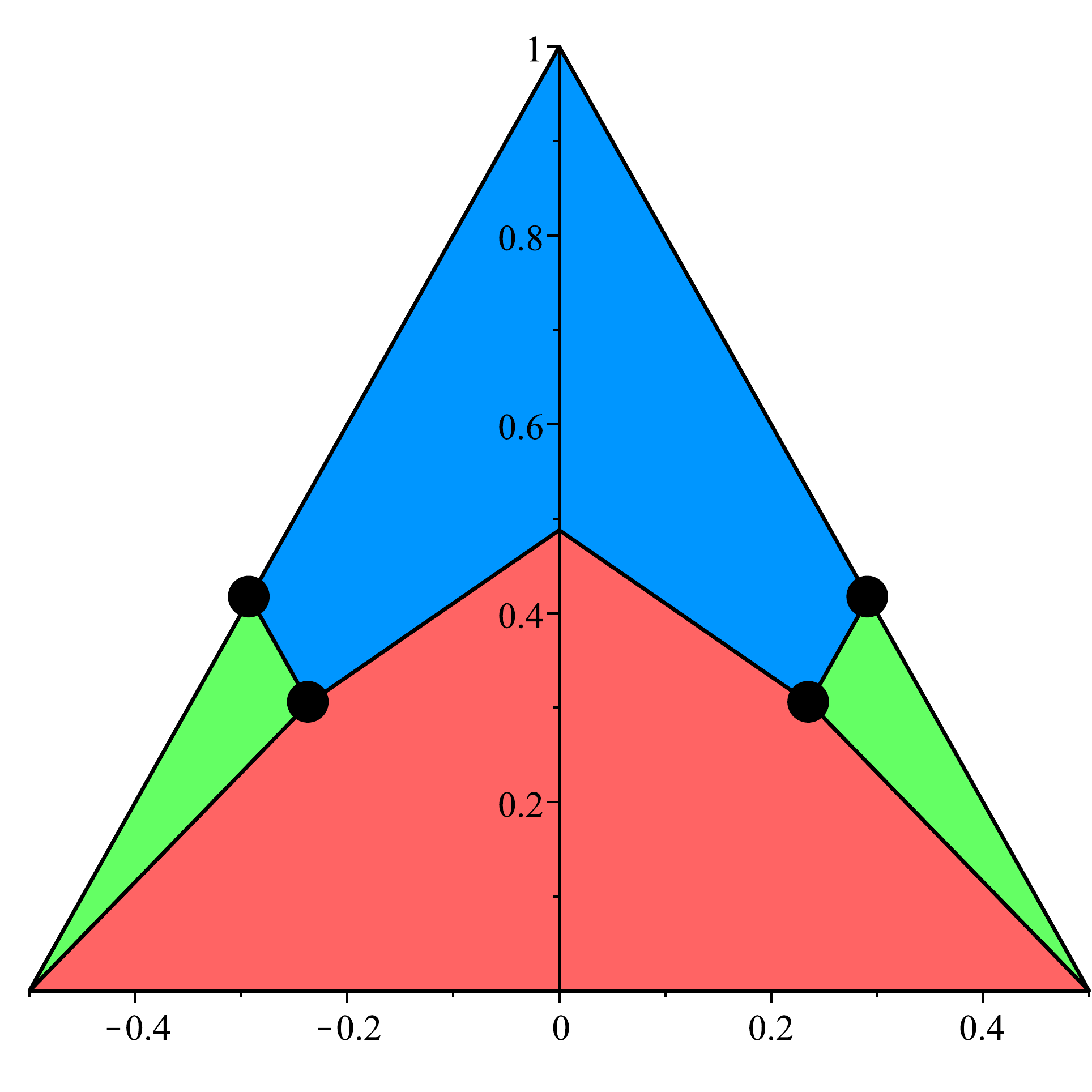} }
\subfloat[$z_{H} = 9$]{
\includegraphics[scale=0.2]{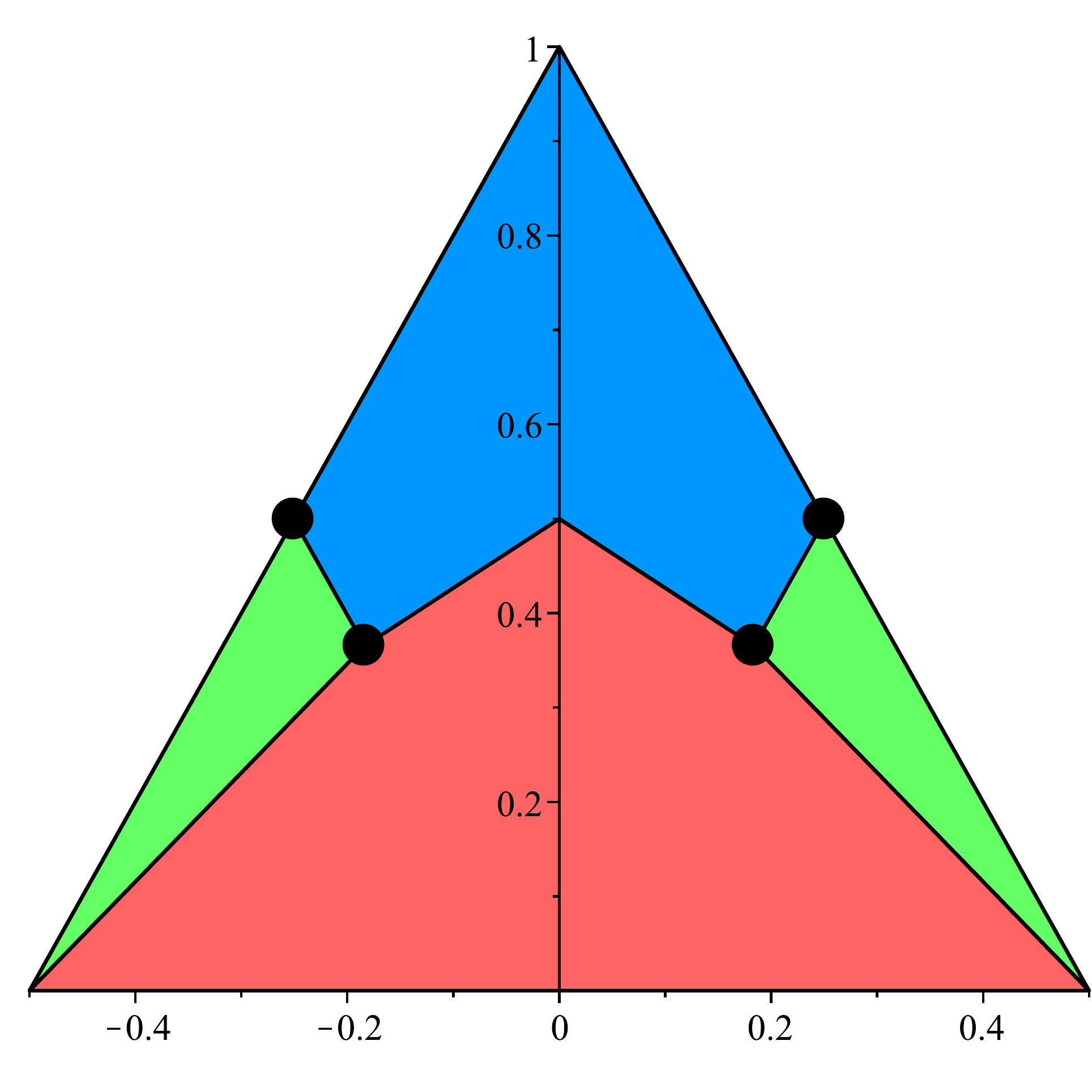} }\caption{The phase
diagrams of the holographic entanglement entropy in the Schwarzschild-AdS
black hole with different horizons. i.e. different temperatures. The critical
points $\left(  \pm x_{t}, a_{t} \right)  $ and $\left(  \pm x_{c}, a_{c}
\right)  $ change with the temperatures. (a,b,c,d) The holographic
entanglement entropy $S_{EE}^{\mathcal{A}}$ for the entangled region
$\mathcal{A}$. (e,f,g,h) The holographic entanglement entropy $S_{EE}%
^{\mathcal{A}^{c}}$ for its supplementary region $\mathcal{A}^{c}$. For the
large horizon or low temperature, the phase diagrams of $S_{EE}^{\mathcal{A}}$
and $S_{EE}^{\mathcal{A}^{c}}$ approach to each other and become the same as
the phase diagram in the pure AdS bulk spacetime.}%
\label{DifferentZh}%
\end{figure}

The phase diagram of the HEE for the region\ $\mathcal{A}$ in the black hole
case has the similar structure as which in the pure AdS case, but with
different critical points $\left(  \pm x_{t},a_{t}\right)  $ and $\left(  \pm
x_{c},a_{c}\right)  $. Fig.\ref{DifferentZh}(a,b,c,d) shows the critical
points and the phase boundaries in the phase diagrams of the HEE
$S_{EE}^{\mathcal{A}}$ at the different temperatures. In the low temperature
limit, i.e. the large $z_{H}$, the phase diagram in the black hole case is
asymptotic to the phase diagram in the pure AdS case as expected. While in the
high temperature limit, i.e. the small $z_{H}$, both the sky and the rainbow
phases shrink to zero, only the sunset phase survives.

The behaviors of the HEE in these two limits are easy to understand from the
viewpoint of the holographic principle. In the low temperature limit, the
horizon is far away from the conformal boundary $\partial\mathcal{M}$ at $z=0$
where the BQFT lives, so that the horizon hardly affects the shape of the
minimal surface. In addition, the Bekenstein-Hawking entropy density in
Eq.(\ref{T and S}) approaches to zero in this limit and can be neglected.
Therefore, the system in the low temperature limit is the same as that in the
pure AdS spacetime.

On the other hand, in the high temperature limit, the horizon is very close to
the conformal boundary so that the rainbow phase is impossible. In addition,
the Bekenstein-Hawking entropy density is divergent in this limit so that the
sky phase, which includes the horizon, is disfavored. Therefore, only the
sunset phase exists in the high temperature limit.

\begin{figure}[t]
\centerline{\includegraphics[scale=0.55]{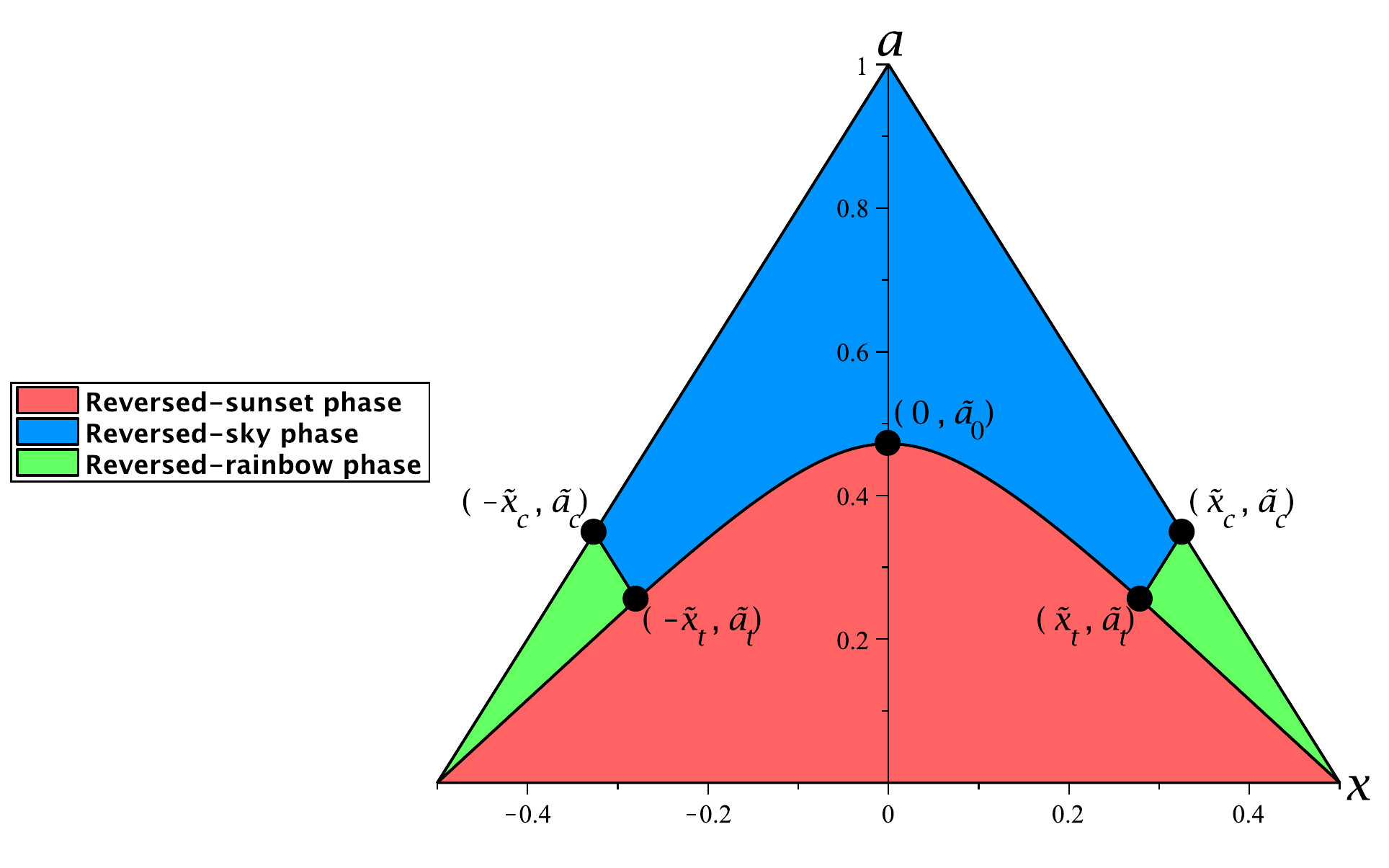}} \caption{Phase
Diagram of the holographic entanglement entropy $S_{EE}^{\mathcal{A}^{c}}$ for
a entangled region $\mathcal{A}$ in the Schwarzschild-AdS blackhole bulk
spacetime with the horizon $zH= 1.5$.}%
\label{AdsZh=1p5CPD}%
\end{figure}

The HEE of\ $\mathcal{A}^{c}$ has the similar behavior with the critical
points $\left(  \pm\tilde{x}_{t},\tilde{a}_{t}\right)  $ and $\left(
\pm\tilde{x}_{c},\tilde{a}_{c}\right)  $. The phase diagram for the HEE of
$\mathcal{A}^{c}$ is shown in Fig.\ref{AdsZh=1p5CPD}. At the middle $x=0$,
there is a critical value $\tilde{a}_{0}$. For $a<\tilde{a}_{0}$, the
entanglement wedge $\mathcal{R}_{\mathcal{A}^{c}}$ takes the shape of the
reversed-sunset; while for $a>\tilde{a}_{0}$, $\mathcal{E}_{\mathcal{A}^{c}}$
breaks into two parts plus the horizon and the entanglement wedge
$\mathcal{R}_{\mathcal{A}^{c}}$ takes the shape of the reversed-sky. When $x$
is away from the middle at $x=0$, the critical value decreases until it
reaches the triple critical points at $\left(  \pm\tilde{x}_{t},\tilde{a}%
_{t}\right)  $ where a new phase, in which the entanglement wedge
$\mathcal{R}_{\mathcal{A}^{c}}$ takes the shape of the reversed-rainbow,
emerges due to the effect of the boundary $\mathcal{Q}$. When $\left\vert
x\right\vert $ is beyond another critical points at $\left(  \pm\tilde{x}%
_{c},\tilde{a}_{c}\right)  $, the reversed-sky phase disappears, and the
reversed-sunset phase and the reversed-rainbow phase compete until $x$ reaches
the boundaries at $x=\pm l/2$.

However, the critical points $\left(  \pm\tilde{x}_{t}, \tilde{a}_{t} \right)
$ and $\left(  \pm\tilde{x}_{c}, \tilde{a}_{c} \right)  $ for $\mathcal{A}%
^{c}$ shift with the temperature in the opposite way of $\left(  \pm x_{t},
a_{t} \right)  $ and $\left(  \pm x_{c}, a_{c} \right)  $ by a similar
argument for $\mathcal{A}$. In the low temperature limit with $z_{H}%
\rightarrow\infty$, the system is the same as that in the pure AdS spacetime.
While in the high temperature limit with $z_{H}\rightarrow0$, only the
reversed-sky phase exists. The critical points and the phase boundaries in the
phase diagrams of the HEE $S_{EE}^{\mathcal{A}^{c}}$ at the different
temperatures are shown in Fig.\ref{DifferentZh}(e,f,g,h).

\begin{figure}[t]
\centerline{\includegraphics[scale=0.45]{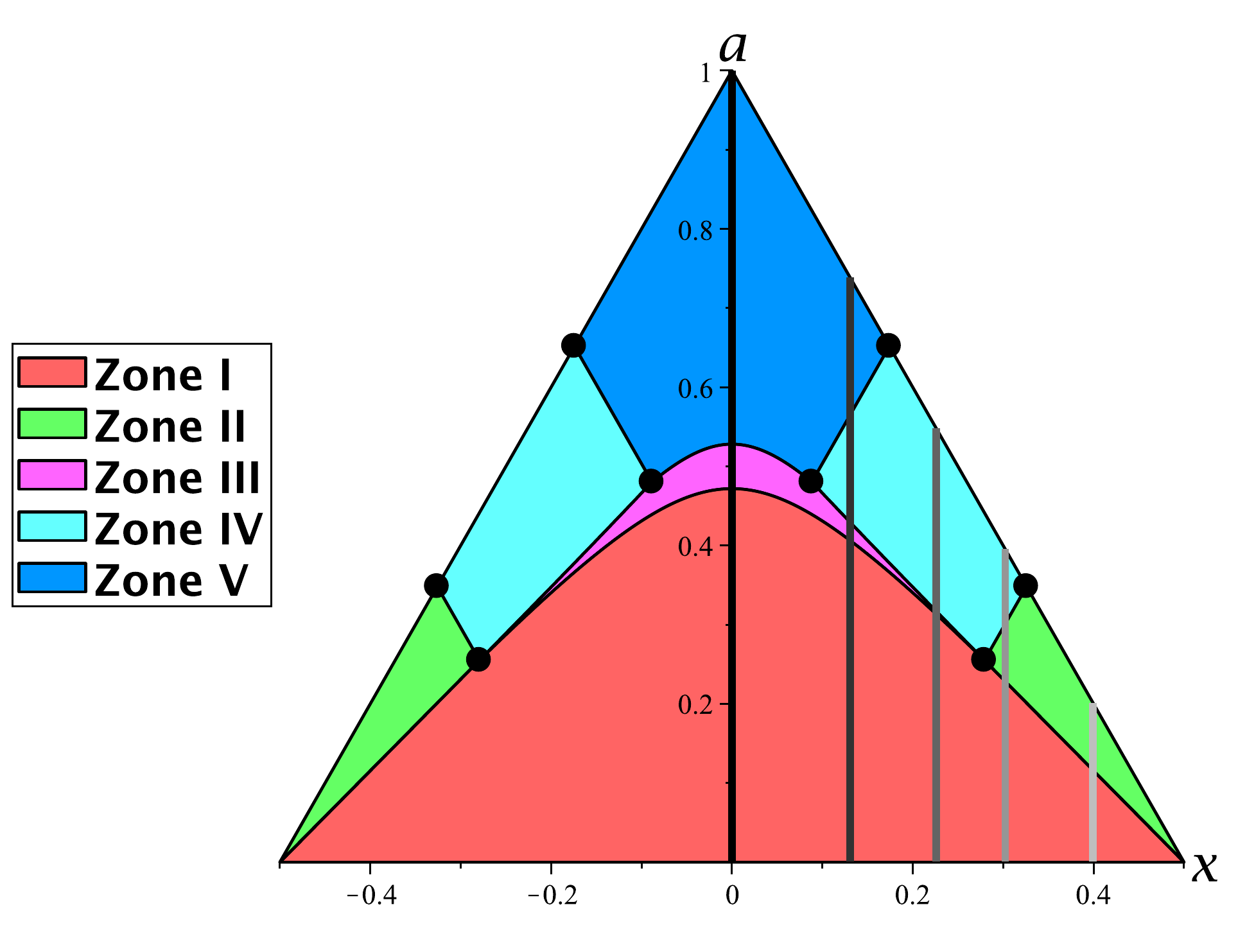}} . \caption{The
phase diagrams of the holographic entanglement entropy $S_{EE}^{\mathcal{A}}$
and $S_{EE}^{\mathcal{A}^{c}}$ for the the entangled region $\mathcal{A}$ and
its complementary $\mathcal{A}^{c}$ are combined together. There are five
zones, marked with different colors, that are associated to the different
phases. The vertical lines represent the different phase transition tracks at
the different location $x=0, 0.13, 0.23, 0.30, 0.40$}%
\label{AdSZh=1p5TPD}%
\end{figure}

\subsection{Entanglement Plateau}

It was conjectured that the HEE satisfies the Araki-Lieb inequality
\begin{equation}
\left\vert \Delta S_{EE}^{\mathcal{A}}\right\vert =\left\vert S_{EE}%
^{\mathcal{A}}-S_{EE}^{\mathcal{A}^{c}}\right\vert \leq S_{BH},
\end{equation}
in the holographic BQFT. To show that, we explore the HEE for $\mathcal{A}$
and $\mathcal{A}^{c}$ in more details by plotting their phase diagrams
together in the Fig.\ref{AdSZh=1p5TPD}. In the phase diagram, there are five
zones as marked in the plot. The associated phase in each zone is listed in
the Table \ref{Stable}.

\begin{table}[ptb]
\begin{center}%
\begin{tabular}
[c]{|c|c|c|c|}\hline
Zone & $\mathcal{A}$ & $\mathcal{A}^{c}$ & $\Delta S_{EE}^{\mathcal{A}}%
$\\\hline
I & Sunset & Reversed-sunset & $-S_{BH}$\\\hline
II & Rainbow & Reversed-rainbow & $-S_{BH}$\\\hline
III & Sunset & Reversed-sky & $\left(  -S_{BH},S_{BH}\right)  $\\\hline
IV & Rainbow & Reversed-sky & $\left(  -S_{BH},S_{BH}\right)  $\\\hline
V & Sky & Reversed-sky & $+S_{BH}$\\\hline
\end{tabular}
\end{center}
\caption{The shapes of $\mathcal{R}_{\mathcal{A}}$ and $\mathcal{R}%
_{\mathcal{A}^{c}}$ in the different zones of the phase diagram in
Fig.\ref{AdSZh=1p5TPD}. The values of $\Delta S_{EE}^{\mathcal{A}}%
=S_{EE}^{\mathcal{A}}-S_{EE}^{\mathcal{A}^{c}}$ for different zones are listed
at the right column of the table.}%
\label{Stable}%
\end{table}

\begin{figure}[t]
\hskip -1 cm \subfloat[3d entanglement plateau]{
\includegraphics[width=10cm]{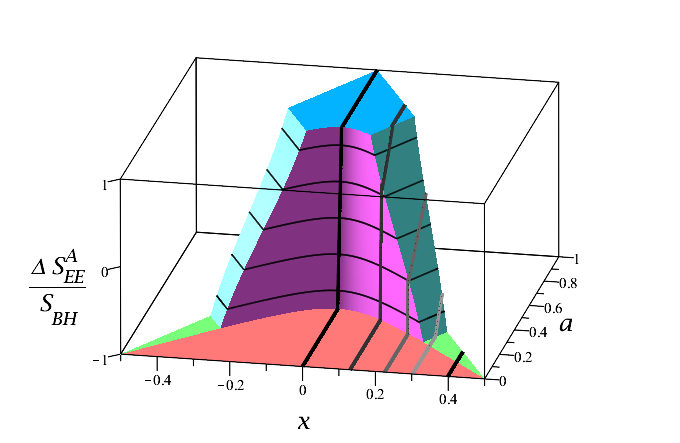} } \hskip -1 cm
\subfloat[Entanglement plateau at different $x$]{
\includegraphics[width=7.5cm]{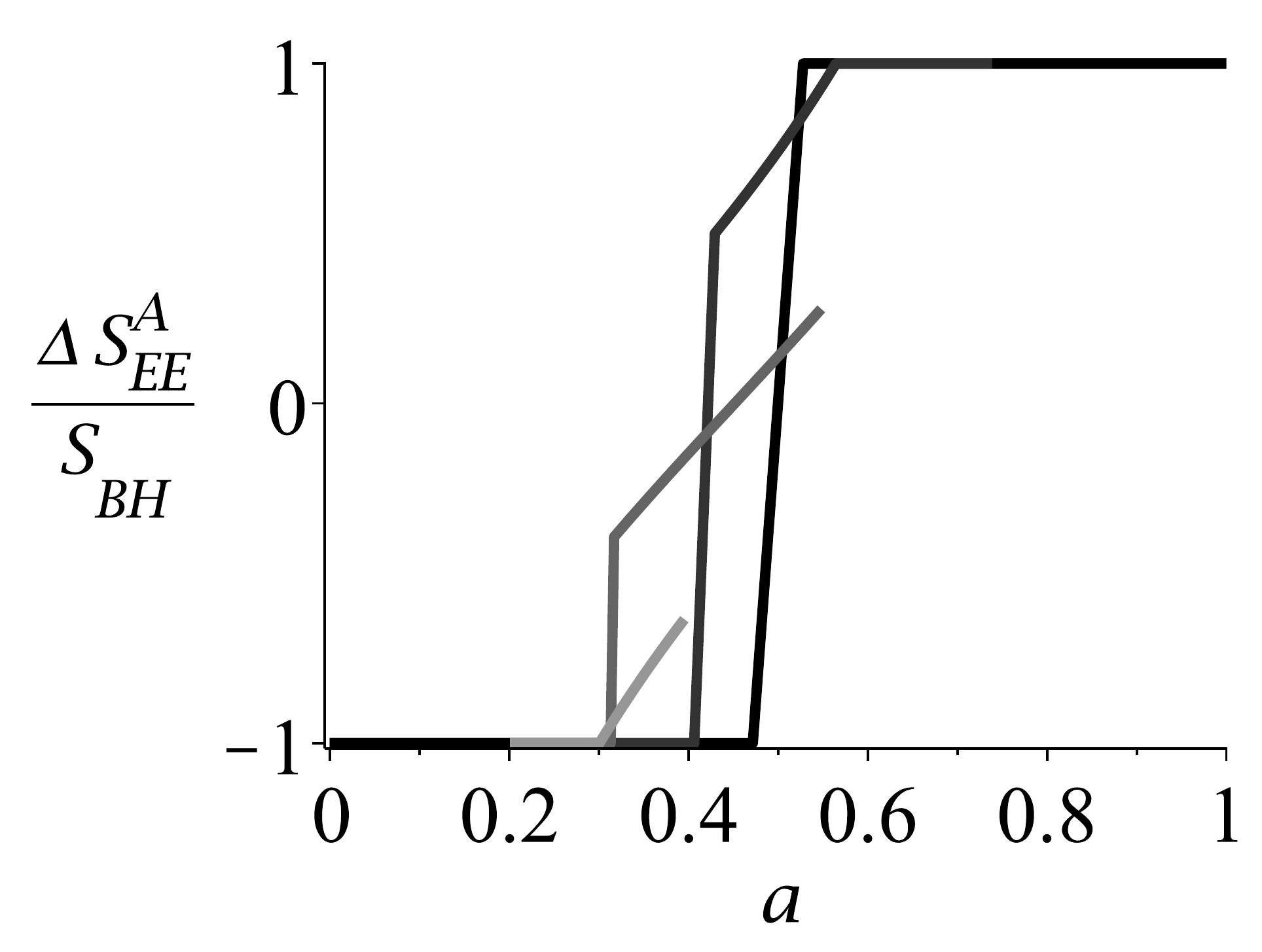} } \caption{(a)
The 3d plot of the holographic entanglement plateau v.s the location $x$ and
the size $a$ of the entangled region $\mathcal{A}$. The different phase
transition tracks at the different location $x=0, 0.13, 0.23, 0.30, 0.40$ are
outlined and plotted in (b).}%
\label{HEEplateaux}%
\end{figure}

We see that $\left\vert \Delta S_{EE}^{\mathcal{A}}\right\vert =S_{BH}$ for
the zones I, II and V. This induces the well known entanglement plateau.
$\Delta S_{EE}^{\mathcal{A}}$ at different location $x$ is plotted in
Fig.\ref{HEEplateaux}(b). For $\left\vert x\right\vert \leq x_{t}$, by
increasing the size $a$, $\Delta S_{EE}^{\mathcal{A}}$ goes through the zones
I-III-V and plots the typical entanglement plateau. For $x_{t}\leq\left\vert
x\right\vert \leq x_{c}$, by increasing the size $a$, $\Delta S_{EE}%
^{\mathcal{A}}$ goes through the zones I-III--IV-V and plots the plateau with
a defected corner. For $x_{c}\leq\left\vert x\right\vert \leq\tilde{x}_{t}$,
by increasing the size $a$, $\Delta S_{EE}^{\mathcal{A}} $ goes through the
zones I-III--IV with the upper plateau disappearing. For $\tilde{x}_{t}%
\leq\left\vert x\right\vert \leq\tilde{x}_{c}$, by increasing the size $a$,
$\Delta S_{EE}^{\mathcal{A}}$ goes through the zones I-II-IV. Finally, for
$\tilde{x}_{c}\leq\left\vert x\right\vert \leq l/2$, by increasing the size
$a$, $\Delta S_{EE}^{\mathcal{A}}$ goes through the zones I-II and always
takes the constant value $-S_{BH}$. The 3d graph of the entanglement plateau
is plotted in Fig.\ref{HEEplateaux}(a).

\section{Summary}

In this paper, we studied the HEE in a $(d+1)$-dimensional holographic BQFT.
We considered two simple solutions for the geometric boundary $\mathcal{Q}$
embedded in the $(d+2)$-dimensional bulk manifolds in the holographic BQFT.
The $AdS_{d+2}$ bulk manifold corresponds to BQFT at the zero temperature, and
the $(d+2)$-dimensional Schwarzschild-AdS black hole bulk manifold corresponds
to BQFT at the finite temperature.

We generalized the Ryu and Takayanagi formula by including the geometric
boundaries and calculated the HEE in both cases. For the pure AdS spacetime,
we found three phases depending on the size and the location of the entangled
region $\mathcal{A}$. We obtained the phase diagram of the HEE in the
holographic BQFT. It is easy to see that the HEE $S_{EE}^{\mathcal{A}}$ for a
region $\mathcal{A}$ is always the same as the HEE $S_{EE}^{\mathcal{A}^{c}}$
for its complementary $\mathcal{A}^{c}$. For the Schwarzschild-AdS black hole
spacetime, we found that the HEE $S_{EE}^{\mathcal{A}}$ is generically not the
same as the HEE $S_{EE}^{\mathcal{A}^{c}}$ due to the homology constraint.
Three new phases were found for the region $\mathcal{A}^{c}$. We obtained the
phase diagrams of the HEE for both $\mathcal{A}$ and $\mathcal{A}^{c}$ and
showed that both of them are asymptotic to that in the pure AdS case in the
low temperature limit as expected.

Furthermore, we verified the Araki-Lieb inequality $\left\vert \Delta
S_{EE}^{\mathcal{A}}\right\vert =\left\vert S_{EE}^{\mathcal{A}}%
-S_{EE}^{\mathcal{A}^{c}}\right\vert \leq S_{BH}$ and obtained the
entanglement plateau by combining the phase diagrams of the HEE for both
$\mathcal{A}$ and $\mathcal{A}^{c}$ together. We plotted the 3d entanglement
plateau v.s the size $a$ and the location $x$ of the entangled region
$\mathcal{A}$.

\section*{Acknowledgements}

We would like to thank Chong-Sun Chu and Rong-Xin Miao for useful discussions.
This work is supported by the Ministry of Science and Technology (MOST
106-2112-M-009 -005 -MY3) and in part by National Center for Theoretical
Science (NCTS), Taiwan.


\begin{thebibliography}{99}                                                                                               %


\bibitem {AL}Huzihiro Araki and Elliott~H Lieb. \newblock Entropy
inequalities. \newblock In \emph{Inequalities}, pages 47--57. Springer, 2002.

\bibitem {9711200}Juan~Martin Maldacena.
\newblock {The Large N limit of superconformal field theories and
supergravity}. \newblock {\em Int. J. Theor. Phys.}, 38:1113--1133, 1999. \newblock [Adv. Theor. Math. Phys.2,231(1998)].

\bibitem {9802150}Edward Witten.
\newblock {Anti-de Sitter space and holography}.
\newblock {\em Adv. Theor. Math. Phys.}, 2:253--291, 1998.

\bibitem {9802109}S.~S. Gubser, Igor~R. Klebanov, and Alexander~M. Polyakov.
\newblock {Gauge theory correlators from noncritical string theory}.
\newblock {\em Phys. Lett.}, B428:105--114, 1998.

\bibitem {0603001}Shinsei Ryu and Tadashi Takayanagi.
\newblock {Holographic derivation of entanglement entropy from AdS/CFT}.
\newblock {\em Phys. Rev. Lett.}, 96:181602, 2006.

\bibitem {0605073}Shinsei Ryu and Tadashi Takayanagi.
\newblock {Aspects of Holographic Entanglement Entropy}. \newblock {\em JHEP},
08:045, 2006.

\bibitem {0705.0016}Veronika~E. Hubeny, Mukund Rangamani, and Tadashi
Takayanagi. \newblock {A Covariant holographic entanglement entropy proposal}.
\newblock {\em JHEP}, 07:062, 2007.

\bibitem {0606184}Dmitri~V. Fursaev.
\newblock {Proof of the holographic formula for entanglement entropy}.
\newblock {\em JHEP}, 09:018, 2006.

\bibitem {1006.0047}Matthew Headrick.
\newblock {Entanglement Renyi entropies in holographic theories}.
\newblock {\em Phys. Rev.}, D82:126010, 2010.

\bibitem {1102.0440}Horacio Casini, Marina Huerta, and Robert~C. Myers.
\newblock {Towards a derivation of holographic entanglement entropy}.
\newblock {\em JHEP}, 05:036, 2011.

\bibitem {1304.4926}Aitor Lewkowycz and Juan Maldacena.
\newblock {Generalized gravitational entropy}. \newblock {\em JHEP}, 08:090, 2013.

\bibitem {1609.01287}Mukund Rangamani and Tadashi Takayanagi.
\newblock {Holographic Entanglement Entropy}.
\newblock {\em Lect. Notes Phys.}, 931:pp.1--246, 2017.

\bibitem {0704.3719}Matthew Headrick and Tadashi Takayanagi.
\newblock {A Holographic proof of the strong subadditivity of entanglement
entropy}. \newblock {\em Phys. Rev.}, D76:106013, 2007.

\bibitem {0710.2956}Tatsuo Azeyanagi, Tatsuma Nishioka, and Tadashi
Takayanagi.
\newblock {Near Extremal Black Hole Entropy as Entanglement Entropy via
AdS(2)/CFT(1)}. \newblock {\em Phys. Rev.}, D77:064005, 2008.

\bibitem {1305.3182}David~D. Blanco, Horacio Casini, Ling-Yan Hung, and
Robert~C. Myers. \newblock {Relative Entropy and Holography}.
\newblock {\em JHEP}, 08:060, 2013.

\bibitem {1306.4004}Veronika~E. Hubeny, Henry Maxfield, Mukund Rangamani, and
Erik Tonni. \newblock {Holographic entanglement plateaux}.
\newblock {\em JHEP}, 08:092, 2013.

\bibitem {1105.5165}Tadashi Takayanagi. \newblock {Holographic Dual of BCFT}.
\newblock {\em Phys. Rev. Lett.}, 107:101602, 2011.

\bibitem {1205.1573}Masahiro Nozaki, Tadashi Takayanagi, and Tomonori Ugajin.
\newblock {Central Charges for BCFTs and Holography}. \newblock {\em JHEP},
06:066, 2012.

\bibitem {1108.5152}Mitsutoshi Fujita, Tadashi Takayanagi, and Erik Tonni.
\newblock {Aspects of AdS/BCFT}. \newblock {\em JHEP}, 11:043, 2011.

\bibitem {1305.2334}Dmitri~V. Fursaev.
\newblock {Quantum Entanglement on Boundaries}. \newblock {\em JHEP}, 07:119, 2013.

\bibitem {1309.4523}Kristan Jensen and Andy O'Bannon.
\newblock {Holography, Entanglement Entropy, and Conformal Field Theories with
Boundaries or Defects}. \newblock {\em Phys. Rev.}, D88(10):106006, 2013.

\bibitem {1403.6475}John Estes, Kristan Jensen, Andy O'Bannon, Efstratios
Tsatis, and Timm Wrase. \newblock {On Holographic Defect Entropy}.
\newblock {\em JHEP}, 05:084, 2014.

\bibitem {1509.02160}Kristan Jensen and Andy O'Bannon.
\newblock {Constraint on Defect and Boundary Renormalization Group Flows}.
\newblock {\em Phys. Rev. Lett.}, 116(9):091601, 2016.

\bibitem {1601.06418}Dmitri~V. Fursaev and Sergey~N. Solodukhin.
\newblock {Anomalies, entropy and boundaries}. \newblock {\em Phys. Rev.},
D93(8):084021, 2016.

\bibitem {1604.07571}Clement Berthiere and Sergey~N. Solodukhin.
\newblock {Boundary effects in entanglement entropy}.
\newblock {\em Nucl. Phys.}, B910:823--841, 2016.

\bibitem {1702.00566}Amin Faraji~Astaneh and Sergey~N. Solodukhin.
\newblock {Holographic calculation of boundary terms in conformal anomaly}.
\newblock {\em Phys. Lett.}, B769:25--33, 2017.

\bibitem {1708.05080}Amin Faraji~Astaneh and Sergey~N. Solodukhin.
\newblock {Holographic calculation of boundary terms in conformal anomaly}.
\newblock {\em Phys. Lett.}, B769:25--33, 2017.

\bibitem {1701.04275}Rong-Xin Miao, Chong-Sun Chu, and Wu-Zhong Guo.
\newblock {New proposal for a holographic boundary conformal field theory}.
\newblock {\em Phys. Rev.}, D96(4):046005, 2017.

\bibitem {1701.07202}Chong-Sun Chu, Rong-Xin Miao, and Wu-Zhong Guo.
\newblock {On New Proposal for Holographic BCFT}. \newblock {\em JHEP},
04:089, 2017.

\bibitem {1410.7811}Johanna Erdmenger, Mario Flory and Max-Niklas Newrzella.
\newblock {Bending branes for DCFT in two dimensions}. \newblock {\em JHEP},
01:058, 2015.

\bibitem {1511.03666}Johanna Erdmengera, Mario Florya, Carlos Hoyosb,
Max-Niklas Newrzellaa and Jackson M. S. Wu.
\newblock {Entanglement Entropy in a Holographic Kondo Model}.
\newblock {\em Fortsch.Phys.}, 64:109--130, 2016.

\bibitem {1703.04186}Amin Faraji~Astaneh, Clement Berthiere, Dmitri Fursaev,
and Sergey~N. Solodukhin.
\newblock {Holographic calculation of entanglement entropy in the presence of
boundaries}. \newblock {\em Phys. Rev.}, D95(10):106013, 2017.
\end{thebibliography}
\end{document}